\documentclass[12pt]{article}

\usepackage{amsfonts,amssymb,graphics,epsfig,verbatim,bm,latexsym,amsmath,url,amsbsy,
geometry}
\newtheorem{theorem}{Theorem}

\usepackage{setspace}
\usepackage{sectsty}
\usepackage[labelformat=empty]{caption}
\usepackage[authoryear]{natbib}
\bibpunct{(}{)}{;}{a}{,}{,}

\begin{document}
\bibliographystyle{plainnat}

\title{\bf A Simple Interactive Fixed Effects Estimator for Short Panels}
\author{Robert F. Phillips and Benjamin D. Williams \\ George Washington University}
\date{September 2024}

\onehalfspace

\maketitle

\abstract{We study the interactive effects (IE) model as an extension of the conventional additive effects (AE) model. For the AE model, the fixed effects estimator can be obtained by applying least squares to a regression that adds a linear projection of the fixed effect on the explanatory variables \citep{Mundlak1978, Chamberlain1984}. In this paper, we develop a novel estimator---the projection-based IE (PIE) estimator---for the IE model that is based on a similar approach. We show that, for the IE model, fixed effects estimators that have appeared in the literature are not equivalent to our PIE estimator, though both can be expressed as a generalized within estimator. Unlike the fixed effects estimators for the IE model, the PIE estimator is consistent for a fixed number of time periods with no restrictions on serial correlation or conditional heteroskedasticity in the errors. We also derive a statistic  for testing the consistency of the two-way fixed effects estimator in the possible presence of iterative effects. Moreover, although the PIE estimator is the solution to a high-dimensional nonlinear least squares problem, we show that it can be computed by iterating between two steps, both of which have simple analytical solutions. The computational simplicity is an important advantage relative to other strategies that have been proposed for estimating the IE model for short panels. Finally, we compare the finite sample performance of IE estimators through simulations.}

\newpage
\section{Introduction \label{intro}}

The two-way fixed effects (TWFE) estimator is a widely-used estimator for panel data regression models due to its flexibility, robustness, and computational simplicity. Its application is motivated by the assumption that the regression error can be decomposed into an additive effects (AE) structure, consisting of the sum of a time effect, an individual effect, and an idiosyncratic error term. If the model's regressors are strictly exogenous with respect to the idiosyncratic error, then the estimator is consistent as the number of individuals ($n$) grows without any restrictions on serial correlation or conditional heteroskedasticity in the errors. The estimator is also consistent under other asymptotic sequences where both $n$ and the number of time periods ($T$) grow or only $T$ grows, though in these cases the serial correlation in the errors must be restricted \citep{Hansen2007}. 

The TWFE estimator is easy to implement and can be computed using one of several equivalent formulations. It can be computed as the least squares solution where both the individual and time effects are treated as parameters to be estimated.\footnote{This is sometimes referred to as the ``least squares dummy variable'' (LSDV) approach, as it can be implemented by including dummy variables for each individual and time period in a pooled OLS regression.}
  Second, the same estimator can be computed by first projecting the individual effects onto the vector of regressors from all time periods, or the time series average of the regressors, and then obtaining the least squares estimator of the modified model \citep{Mundlak1978, Chamberlain1984}. 
Finally, both are also equivalent to first applying a within transformation to the regressors and then calculating the ordinary least squares (OLS) estimator.

Despite these advantages of the TWFE estimator, it is motivated by a strong functional form assumption that any unobserved individual characteristics that are correlated with the regressors must be time-invariant. In some contexts this is interpreted as an assumption of parallel trends --- individual-specific time trends must be parallel. One alternative specification that is commonly used is to allow for individual-specific linear time trends \citep[see, e.g.,][]{Heckman1989}. 

A specification that relaxes the strong functional form assumption of additive effects and does not rely on individual-specific linear time trends is the interactive effects (IE) model.  In the IE model the individual effect is multiplied by an unknown time-varying coefficient. This specification dates back to at least \cite{Kiefer1980} and has since been extended to allow for multiple time-invariant individual effects, each multiplied by a separate time-varying coefficient.\footnote{Microeconometric models with this error structure (i.e., a factor structure) have also been used by \cite{Goldberger1972}, \cite{Joreskog1975}, \cite{Heckman1987}, among others, but \cite{Kiefer1980} appears to be the first to introduce it explicitly as an extension of the TWFE model.} In the presence of interactive effects, the TWFE estimator is inconsistent if the covariance between an individual effect and at least one regressor is time-varying (see Section 2). 

The fixed effects approach to estimating the AE model, i.e., the least squares solution where both the individual and time effects are treated as \textit{fixed} parameters to be estimated, has been generalized to the IE model by \cite{Kiefer1980}, \cite{Lee1991}, \cite{Ahn2001}, and \cite{Bai2009}, among others. \cite{Kiefer1980}, \cite{Lee1991}, and \cite{Ahn2001} show that their fixed effects approach yields a generalized within estimator that is consistent for a fixed $T$ only under the assumption that the regression error is serially uncorrelated and has constant variance over time, a prohibitively strong assumption in most contexts. However, \cite{Bai2009} shows that, under asymptotics where both  $n$ and $T$ grow, this estimator is consistent with heteroskedastic and weakly serially correlated errors. For this reason we refer to this as a large-$T$ IE estimator. 

Numerous recent studies have found the IE model to be an empirically relevant extension of the AE model \citep[e.g.,][]{Kim2014, Gobillon2016,Totty2017,Juodis2022}. And many of these studies apply Bai's (2009) or other large-$T$ methods. However, many empirical applications of the TWFE estimator, or related difference-in-differences methods, rely on small- to moderate-length panels and often use dependent variables that are highly serially correlated. In a survey of difference-in-differences papers, \cite{Bertrand2004} found an average panel length of 16.5 periods and found that the most commonly used dependent variables were employment and wages, both of which can be highly serially correlated. Small-$T$ methods for the IE model have been proposed by \cite{Hayakawa2012}, \cite{Ahn2013}, \cite{Robertson2015}, \cite{Westerlundetal2019}, \cite{Juodis2022}, \cite{Imbens2021}, and \cite{Callaway2023}. While several of these methods are more general than the method we describe in this paper, ours has the comparative advantage of computational simplicity.  Moreover, we show how it can be used to test for the adequacy of the TWFE estimator.

In this paper, we implement the projection approach for the IE model. This approach has been much less common in the literature on the IE model. \cite{Bai2013} and \cite{Hayakawa2012} are two notable exceptions, and both  focus on a dynamic panel model. We apply the projection approach analogously to \cite{Chamberlain1984}'s well-known result for the AE model. We replace the time-invariant individual effect by a projection on the vector of regressors from all periods. Because the IE model is linear in the individual effects, this is done without loss of generality, unlike in nonlinear models.\footnote{Thus, this is still a fixed effects model, not a random effects model, in that only mild regularity conditions are placed on the distribution of the individual effects and no restrictions on correlation with the regressors are imposed.}  While this projection-based estimator for the IE model is---like the large-$T$ IE estimator---a generalized within estimator, it is not equivalent to the large-$T$ IE estimator produced by the fixed effects approach, in contrast to the AE model. And projecting the time-invariant individual effect onto the regressors has an advantage. Specifically, we show that the estimator it produces is consistent for fixed $T$ given arbitrarily serially correlated and conditionally heteroskedastic errors, provided the regressors are strictly exogenous. We refer to this as the small-$T$ projection-based IE (PIE) estimator. Further, we show that, similar to the large-$T$ IE estimator, it can be computed through a simple iterative scheme where one step involves a generalized within transformation and the other involves an eigenvalue decomposition.

We also propose a test statistic that can be used to test for the consistency of the TWFE estimator.  Although the TWFE estimator is generally motivated assuming AE errors, it is not necessarily inconsistent in the presence of IE errors.  We show that it is only inconsistent given IE errors provided the correlation between an individual effect and at least one regressor is time-varying.   Using this fact, we  provide a large sample joint distribution of our new estimator and the TWFE estimator which we then use  to derive a statistic for testing for the consistency of the TWFE estimator.

The remainder of the paper is organized as follows. In Section~\ref{model} we provide the IE model and derive the bias of the TWFE estimator. In Section~\ref{estimator} we provide the linear projection that is the basis of the estimation approach used in the paper.  We also provide a necessary and sufficient identification condition, as well as a simple necessary identification condition which can be used to check the identification status of an IE model.  Moreover, in Section~\ref{estimator} we define the small-$T$ PIE estimator and derive the large-$n$ asymptotic behavior of the  estimator.  In Section~\ref{test} we use the asymptotic theory developed in Section~\ref{estimator} to derive a statistic for testing the consistency of the TWFE estimator.  The statistic relies on contrasts, similar to \cite{Westerlund2019}. In Section \ref{computation} we describe an algorithm that can be used to compute PIE estimates and discuss the relationship between the small-$ T $ PIE and large-$T$ IE estimators.  Finally, in Section~\ref{monte} we discuss computation of the small-$T$ PIE estimator and study its finite sample behavior in two Monte Carlo studies. 

\section{The IE model and bias of the TWFE estimator\label{model}}

The model examined in this paper is 
\begin{equation}\label{outcome}
\boldsymbol{y}_{i}=\boldsymbol{X}_{i}\boldsymbol{\beta}_0 + \boldsymbol{e}_{i}, \qquad (i=1,\ldots,n).
\end{equation}
In this model $\boldsymbol{X}_i$ is a $T \times K$ matrix with the  vector of explanatory variables $\boldsymbol{x}_{it}^{\prime}$ in its $t$th row ($ t=1,\ldots,T $). Moreover, $\boldsymbol{e}_i =  (e_{i1},\ldots,e_{iT})^{\prime}$ is a vector of errors, while $\boldsymbol{y}_i =  (y_{i1},\ldots,y_{iT})^{\prime}$ is a vector of observed values of the dependent variable for the $ i $th cross-sectional unit, which we refer to as an  ``individual''.  Assuming randomly sampled individuals, our primary interest is in estimating the $ K \times 1 $ vector of parameters $ \boldsymbol{\beta}_0 $.

To that end, researchers often assume the regression errors are generated by the AE model:
\begin{equation*}
e_{i,t} = \delta_{0t} + \kappa_{0i} + \epsilon_{it} \qquad (i=1,\ldots,n, \; t=1,\ldots,T).
\end{equation*}
Given the AE model for the errors, if $E(\epsilon_{it}|\boldsymbol{x}_{it},\delta_{0t},\kappa_{0i}) = 0$,  then $ \boldsymbol{\beta}_0 $  can be consistently estimated with the TWFE estimator.  

\sloppy
Consistent estimation of $ \boldsymbol{\beta}_0 $ becomes more complicated when the errors are generated by an IE process.  Specifically, suppose the errors are generated according to 
\begin{equation} \label{ie}
e_{it} = \delta_{0t} + \boldsymbol{\phi}_{0t}^{\prime}\boldsymbol{\widetilde{\eta}}_i + \epsilon_{it} \qquad (i=1,\ldots,n, \; t=1,\ldots,T).
\end{equation}
where $ \boldsymbol{\widetilde{\eta}}_i = (\widetilde{\eta}_{i1},\ldots,\widetilde{\eta}_{iq})^{\prime}$ is a vector of $ q $ unobserved time-invariant characteristics which are interacted with  time-varying parameters in the $ q \times 1 $ vector $ \boldsymbol{\phi}_{0t}=(\phi_{0t1},\ldots,\phi_{0tq})'$. The entries in $ \boldsymbol{\widetilde{\eta}}_i $ are referred to as common factors while those in $ \boldsymbol{\phi}_{0t} $ are factor loadings.\footnote{This terminology is sometimes reversed in the IE literature. The terms originate in the psychometrics literature on factor models, where typically the common factors are treated as random variables and the factor loadings as parameters.} Analogous to the strict exogeneity often assumed when the AE model is employed, we assume $E(\epsilon_{it}|\boldsymbol{x}_{it},\delta_{0t},\boldsymbol{\phi}_{0t}^{\prime}\boldsymbol{\widetilde{\eta}}_i) = 0$.  Moreover, because the error term, $ e_{it} $, includes the time effect, $ \delta_{0t} $, the generality of the specification is not affected by imposing the restriction $ E(\boldsymbol{\widetilde{\eta}}_i)  = \boldsymbol{0} $. The conventional AE model with time and individual effects is a special case of this model with $ \boldsymbol{\phi}_{0t} = \boldsymbol{\phi}_{0} $ ($ t=1,\ldots,T $), in which case $\kappa_{0i} = \boldsymbol{\phi}_{0}^{\prime}\boldsymbol{\widetilde{\eta}}_i $.

\fussy
In light of the IE specification in (\ref{ie}), the model in (\ref{outcome}) can be written as
\begin{equation}\label{model1}
	\boldsymbol{y}_i = \boldsymbol{{X}}_i\boldsymbol{\beta}_0 + \boldsymbol{\delta}_{0} + \boldsymbol{\Phi}_0 \boldsymbol{\widetilde{\eta}}_i + \boldsymbol{\epsilon}_i \qquad (i=1,\ldots,n),
\end{equation}
where $\boldsymbol{\delta}_{0} = (\delta_{01},\ldots,\delta_{0T})^{\prime}$; $\boldsymbol{\epsilon}_i =  (\epsilon_{i1},\ldots,\epsilon_{iT})^{\prime}$; and the $T \times q$ matrix $\boldsymbol{\Phi}_0$ is formed by stacking $\boldsymbol{\phi}_{0t}^{\prime}$ $(t=1,\ldots,T)$. Although this model  only simplifies to the conventional AE model when the factor loadings, $ \boldsymbol{\phi}_{0t}$, are time-invariant, the TWFE estimator remains consistent for estimating $ \boldsymbol{\beta}_0 $ under less restrictive conditions.  For the TWFE estimator to be biased, at least one of the common factors in $ \boldsymbol{\widetilde{\eta}}_i $ must have both a time-varying factor loading and a time-varying covariance with at least one of the explanatory variables. 

To see this, consider the linear projection of $\boldsymbol{x}_{it}$ onto the common factors:
\begin{equation} \label{x_ie}
\boldsymbol{x}_{it} = \boldsymbol{\delta}_{0t}^x + \boldsymbol{\Phi}_{0t}^{x \prime}\boldsymbol{\widetilde{\eta}}_i + \boldsymbol{\epsilon}_{it}^x \qquad (t=1,\ldots,T).
\end{equation}
This decomposition always exists, with $E(\boldsymbol{\widetilde{\eta}}_i \boldsymbol{\epsilon}_{it}^{x \prime})=\boldsymbol{0}$, regardless of whether the factor loadings, $\boldsymbol{\Phi}_{0t}^x$, have a ``structural'' interpretation. The factor loadings are given by $\boldsymbol{\Phi}_{0t}^x=\text{Var}(\boldsymbol{\widetilde{\eta}}_i)^{-1}\text{Cov}(\boldsymbol{\widetilde{\eta}}_i,\boldsymbol{x}_{it})$.  Note that $ \boldsymbol{\Phi}_{0t}^x $ does not vary with $ t $ if $ \text{Cov}(\boldsymbol{\widetilde{\eta}}_i,\boldsymbol{x}_{it}) $ does not vary with $ t $.
Next, define $\ddot{\boldsymbol{x}}_{it}:=\boldsymbol{x}_{it}-\bar{\boldsymbol{x}}_{i\cdot}-(\bar{\boldsymbol{x}}_{\cdot t}-\bar{\boldsymbol{x}})$ ($t=1,\ldots, T$), where $\bar{\boldsymbol{x}}_{i\cdot}=(1/T)\sum_t \boldsymbol{x}_{it}$, $\bar{\boldsymbol{x}}_{\cdot t}=(1/n)\sum_i \boldsymbol{x}_{it}$, and $\bar{\boldsymbol{x}}=(1/(nT))\sum_{i,t} \boldsymbol{x}_{it}$. Moreover, let $\ddot{y}_{it}$  denote the same within transformation of $y_{it}$ ($ t=1,\ldots,T $). It is well-known that the TWFE estimator is equal to $\left(\sum_{i,t} \ddot{\boldsymbol{x}}_{it}\ddot{\boldsymbol{x}}_{it}'\right)^{-1}\sum_{i,t} \ddot{\boldsymbol{x}}_{it}\ddot{y}_{it}$.   The large-$n$ bias of this estimator (for fixed $T$), under the model given by equations (\ref{outcome}) and (\ref{ie}) with $E(\boldsymbol{x}_{it}\epsilon_{is})=\boldsymbol{0}$ for all $s$ and $t$, is
\begin{equation} \label{bias}
\left(\frac{1}{T}\sum_{t}E\left(\ddot{\boldsymbol{x}}_{it}\ddot{\boldsymbol{x}}_{it}\right)'\right)^{-1}\frac{1}{T}\sum_t\left(\boldsymbol{\Phi}_{0t}^x-\bar{\boldsymbol{\Phi}}_{0}^x\right)' E\left(\boldsymbol{\widetilde{\eta}}_i\boldsymbol{\widetilde{\eta}}_i'\right)\left(\boldsymbol{\phi}_{0t}-\bar{\boldsymbol{\phi}}_{0}\right).
\end{equation}
where $\bar{\boldsymbol{\Phi}}_{0}^x = (1/T)\sum_t {\boldsymbol{\Phi}}_{0t}^x$ and $\bar{\boldsymbol{\phi}}_{0} = (1/T)\sum_t {\boldsymbol{\phi}}_{0t}$. This expression makes clear that the TWFE estimator is consistent if the factor loadings in the error specification are time-invariant {\em or} if the factor loadings for $\boldsymbol{x}_{it} $ are time-invariant. The latter occurs if the covariances between the entries in $\boldsymbol{x}_{it}$ and $\boldsymbol{\widetilde{\eta}}_i$ are not time-varying. However, if both sets of factor loadings are time-varying, then the TWFE estimator will be inconsistent for a fixed $T$.\footnote{This expression also suggests a milder condition for consistency of the TWFE estimator as both $n$ and $T$ grow to infinity.}

The following example illustrates the inconsistency of the difference-in-differences estimator in the absence of parallel trends. Non-parallel trends can be captured with the IE error model.

\subsubsection*{Example 1.} Suppose  $T=2$, $K=1$, and $q=1$. Also, let $x_{i1} =0$ for all $i$ and $x_{i2}\in\{0,1\}$. Then
\begin{equation*}
\begin{split}
y_{i1} & = \delta_{01} + \phi_{01}\widetilde{\eta}_i + \epsilon_{i1} \\
y_{i2} & = \beta_0 x_{i2} + \delta_{02} + \phi_{02}\widetilde{\eta}_i + \epsilon_{i2}.
\end{split}
\end{equation*}
We refer to this as a ``treatment effect'' model. The TWFE estimator for $\beta_0$ is the difference-in-differences estimator $\widehat{E}_n(y_{i2}-y_{i1}\mid  x_{i2}=1)-\widehat{E}_n(y_{i2}-y_{i1}\mid  x_{i2}=0)$. This estimator is biased if $\phi_{02}\neq \phi_{01}$ and $x_{i2}$ is correlated with $\widetilde{\eta}_i$.  Indeed, the bias can be written as $(\phi_{02}-\phi_{01})\left[E(\widetilde{\eta}_i\mid x_{i2}=1)-E(\widetilde{\eta}_i\mid x_{i2}=0)\right]$. The covariance between $x_{it}$ and $\widetilde{\eta}_i$ is time-varying by construction except in the trivial case where it is $0$ for all $t$.

\section{The small-$T$ PIE estimator \label{estimator}}

\subsection{A linear projection}
Our objective is to estimate $\boldsymbol{\beta}_0$, but estimating $\boldsymbol{\beta}_0$ requires estimating time-varying parameters, such as those in  $ \boldsymbol{\Phi}_0 $.  However, the time-varying parameters cannot be uniquely identified due to a fundamental indeterminancy in the absence of a scaling restriction.  To see this, note that the interactive effect in period $t$, $ \boldsymbol{\phi}_{0t}^{\prime}\boldsymbol{\widetilde{\eta}}_i $, is indistinguishable from $ \boldsymbol{\phi}_{0t}^{\ast \prime}\boldsymbol{\widetilde{\eta}}_i^{\ast} $, for $ \boldsymbol{\phi}_{0t}^{\ast} = \boldsymbol{C}'\boldsymbol{\phi}_{0t} $ and $ \boldsymbol{\widetilde{\eta}}_i^{\ast} =  \boldsymbol{C}^{-1}\boldsymbol{\widetilde{\eta}}_i$ for any nonsingular $q \times q$ matrix $\boldsymbol{C}$. 

The indeterminancy can be resolved as follows. Let $ \boldsymbol{\Phi}_{01} $ and $ \boldsymbol{\Phi}_{02} $ be the first $ q $ rows and last $ T-q $ rows of the $ T \times q $ matrix $ \boldsymbol{\Phi}_{0} $, respectively, assuming that $ 1 \leq q < T $. If $ \boldsymbol{\Phi}_{01} $ is nonsingular, then we can define $ \boldsymbol{\eta}_{i} := \boldsymbol{\Phi}_{01}\boldsymbol{\widetilde{\eta}}_{i} $, $ \boldsymbol{\Lambda}_{02} := \boldsymbol{\Phi}_{02}\boldsymbol{\Phi}_{01}^{-1} $, and  
\begin{equation*}
\boldsymbol{\Lambda}_0  :=  \left( \begin{array}{c}
\boldsymbol{I}_{q} \\
\boldsymbol{\Lambda}_{02}
\end{array} \right) .
\end{equation*}
\sloppy
Because $  \boldsymbol{\Lambda}_0\boldsymbol{\eta}_i = \boldsymbol{\Phi}_{0}\boldsymbol{\widetilde{\eta}}_{i}$, the model in (\ref{model1}) can be rewritten as 
\begin{equation*}
\boldsymbol{y}_i = \boldsymbol{\delta}_{0}+\boldsymbol{{X}}_i\boldsymbol{\beta}_0 +  \boldsymbol{\Lambda}_0 \boldsymbol{\eta}_i + \boldsymbol{\epsilon}_i \; \; \; \; \; \; \; \; \; \; \; \; (i=1,\ldots,n).
\end{equation*}

When the unobserved effects in $\boldsymbol{\eta}_i$ are  uncorrelated with the regressors, the time effects in $ \boldsymbol{\delta}_0 $ and the regression parameters in $ \boldsymbol{\beta}_0 $ can be estimated with ordinary least squares.  The more complicated case is when the unobserved effects are correlated with the regressors. If $T$ is small relative to $n$, we can control for this correlation using Mundlak-Chamberlain linear projections \citep{Mundlak1978, Chamberlain1984}. In particular, let $\boldsymbol{z}_i$ be an $m \times 1$ vector consisting of all the distinct entries in $(\boldsymbol{x}_{i1}',\ldots, \boldsymbol{x}_{iT}')'$.\footnote{If $E(\boldsymbol{z}_i\boldsymbol{z}_i')$ is not full rank we remove additional components until it is full rank.} Then the linear projection of the $ j $th entry in $ \boldsymbol{\eta}_i $ on 1 and $\boldsymbol{z}_i$ is
\begin{equation} \label{eta_lp}
\eta_{ij} = \mu_{0j} + \boldsymbol{z}_i^{\prime}\boldsymbol{\theta}_{0j} + a_{ij}  \qquad (j=1,\ldots,q),
\end{equation}
where $ \boldsymbol{\theta}_{0j} = \text{Var}(\boldsymbol{z}_i)^{-1}\text{Cov}(\boldsymbol{z}_i,\eta_{ij}) = \text{Var}(\boldsymbol{z}_i)^{-1}E(\boldsymbol{z}_i\eta_{ij}) $ and $ \mu_{0j} = E\left( \eta_{ij}\right) - E\left( \boldsymbol{z}_{i}' \right) \boldsymbol{\theta}_{0j} = - E\left( \boldsymbol{z}_{i}' \right) \boldsymbol{\theta}_{0j} $.  The formulas for $ \boldsymbol{\theta}_{0j} $ and $ \mu_{0j} $ exploit the fact that $ E\left( \widetilde{\eta}_{ij}\right) = 0  $ implies $ E\left( \eta_{ij}\right) = 0  $.  Note that the linear projection residual $a_{ij}$  has, by construction, mean zero and is uncorrelated with the entries in $\boldsymbol{z}_i$. Using these linear projections, we obtain the augmented nonlinear regression model
\begin{equation} \label{aug}
\boldsymbol{y}_i = \boldsymbol{\delta}_{0}^{\ast}+\boldsymbol{{X}}_i\boldsymbol{\beta}_0 +  \boldsymbol{\Lambda}_0\boldsymbol{Z}_i\boldsymbol{\theta}_0 + \boldsymbol{u}_i \qquad (i=1,\ldots,n),
\end{equation}
where  $ \boldsymbol{Z}_i := \boldsymbol{I}_q \otimes \boldsymbol{z}_i^{\prime}    $;   $ \boldsymbol{\theta}_0 := \text{vec}(\boldsymbol{\Theta}_0) $ and $\boldsymbol{\Theta}_0:= ( \boldsymbol{\theta}_{01},\ldots, \boldsymbol{\theta}_{0q}) $; $ \boldsymbol{\delta}_0^{\ast} :=  \boldsymbol{\delta}_0 + \boldsymbol{\Lambda}_0 \boldsymbol{\mu}_0$, where $ \boldsymbol{\mu}_0 := (\mu_{01},\ldots,\mu_{0q})' $; and, finally, $ \boldsymbol{u}_{i} = \boldsymbol{\Lambda}_0\boldsymbol{a}_i + \boldsymbol{\epsilon}_{i} $, with $ \boldsymbol{a}_i =(a_{i1},\ldots, a_{iq})^{\prime} $.

Finally, because we are not interested in estimating  $ \boldsymbol{\delta}_0^{\ast} $, we demean the variables.  Specifically, set $ \dot{\boldsymbol{X}}_i := \boldsymbol{X}_i - \overline{\boldsymbol{X}} $, where $ \overline{\boldsymbol{X}} := (1/n)\sum_{i}\boldsymbol{X}_i $, and let $ \dot{\boldsymbol{y}}_i $, $ \dot{\boldsymbol{Z}}_i $, and $ \dot{\boldsymbol{u}}_i $ be defined analogously.  Then 
\begin{equation} \label{aug_demeaned}
	\dot{\boldsymbol{y}}_i = \dot{\boldsymbol{X}}_i\boldsymbol{\beta}_0 +  \boldsymbol{\Lambda}_0\dot{\boldsymbol{Z}}_i\boldsymbol{\theta}_0+ \dot{\boldsymbol{u}}_i \qquad (i=1,\ldots,n).
\end{equation}

\sloppy
Note that if $q=1$ and $\boldsymbol{\Lambda}_0 =  \boldsymbol{\iota}_T  $, where $ \boldsymbol{\iota}_T$ is a $ T \times 1 $ vector of ones,  then $\boldsymbol{\gamma}_0 := (\boldsymbol{\beta}_0', \boldsymbol{\theta}_0')' $ can be estimated with $\left(\sum_i \dot{\boldsymbol{W}}_i '\dot{\boldsymbol{W}}_i\right)^{-1} \sum_i\dot{\boldsymbol{W}}_i \dot{\boldsymbol{y}}_i$, for $ \dot{\boldsymbol{W}}_i = (\dot{\boldsymbol{X}}, \, \boldsymbol{\iota}_T \dot{\boldsymbol{z}}_i')  $. \cite{Mundlak1978} showed that the subvector of this vector corresponding to $\boldsymbol{\beta}_0 $ is identical to the TWFE estimator for $\boldsymbol{\beta}_0$.  Moreover, upon applying the Frisch–Waugh–Lovell (FWL) theorem, we can express the TWFE estimator of $\boldsymbol{\beta}_0$ as
\begin{equation} \label{TWFE}
	\widehat{\boldsymbol{\beta}}_{FE}  =  \left( \sum_{i}\dot{\boldsymbol{X}}_i' \boldsymbol{Q} \dot{\boldsymbol{X}}_i \right)^{-1} \sum_{i}\dot{\boldsymbol{X}}_i' \boldsymbol{Q}\dot{\boldsymbol{y}}_i ,
\end{equation}  
where $ \boldsymbol{Q} := \boldsymbol{I}_{T} -\boldsymbol{\iota}_{T}\left( \boldsymbol{\iota}_{T}'\boldsymbol{\iota}_{T}\right)^{-1} \boldsymbol{\iota}_{T}^{\prime} $ is the within transformation matrix.  

\fussy
\subsection{Identification}
Equation~(\ref{aug_demeaned}) is a nonlinear-in-parameters regression model.  Therefore, if suitable conditions are satisfied, the parameters can, in principle, be estimated via (pooled) nonlinear least squares (NLS). In order to define the NLS objective function, let $ \boldsymbol{\Theta} := (\boldsymbol{\theta}_1,\ldots,\boldsymbol{\theta}_q)$ denote a $m \times q$ matrix, and let $\boldsymbol{\Lambda}' :=(\boldsymbol{I}_q, \boldsymbol{\Lambda}_2') $, where $\boldsymbol{\Lambda}_2$ is a $(T-q) \times q$ matrix.  Moreover, set $\boldsymbol{\theta} := \text{vec}(\boldsymbol{\Theta})$ and   $ \boldsymbol{\lambda} := \text{vec}(\boldsymbol{\Lambda}_2)$. Also, let $ \boldsymbol{\gamma} := ( \boldsymbol{\beta}^{\prime}, \boldsymbol{\theta}^{\prime})^{\prime} $, where $\boldsymbol{\beta}$ is $K \times 1$, and let $ \boldsymbol{\psi} := (\boldsymbol{\gamma}^{\prime},\boldsymbol{\lambda}^{\prime})^{\prime} $ denote an element of the parameter space $ \boldsymbol{\Psi} \subseteq \mathbb{R}^p $ ($ p=K+q(m+T-q) $). Then the NLS  estimator of $\boldsymbol{\psi}_0 = (\boldsymbol{\gamma}_0^{\prime},\boldsymbol{\lambda}_0^{\prime})^{\prime} \in \boldsymbol{\Psi}$  is the minimizer of  
\begin{equation*}
	S_{n}( \boldsymbol{\psi}) := \frac{1}{n}\sum_{i} s(\dot{\boldsymbol{v}}_i,\boldsymbol{\psi}) ,
\end{equation*}
where $ \dot{\boldsymbol{v}}_i :=(\dot{\boldsymbol{z}}_i', \dot{\boldsymbol{a}}_i', \dot{\boldsymbol{\epsilon}}_i')' $, and
$ s(\dot{\boldsymbol{v}}_i,\boldsymbol{\psi}) := ( \dot{\boldsymbol{y}}_i - \dot{\boldsymbol{X}}_i\boldsymbol{\beta} - \boldsymbol{\Lambda}\dot{\boldsymbol{Z}}_i\boldsymbol{\theta}) '( \dot{\boldsymbol{y}}_i - \dot{\boldsymbol{X}}_i\boldsymbol{\beta} - \boldsymbol{\Lambda}\dot{\boldsymbol{Z}}_i\boldsymbol{\theta})/2 $. The PIE estimator of $\boldsymbol{\beta}_0$ is the subvector of this NLS estimator corresponding to $\boldsymbol{\beta}$.

In order for the NLS estimator to consistently estimate $\boldsymbol{\psi}_0$, the latter must be identified. The parameter vector  $\boldsymbol{\psi}_0$ is identified if $S_n(\cdot)$ is uniquely minimized at $\boldsymbol{\psi}_0$ in the limit, as $n \rightarrow \infty$.  To see when this will occur, note that $s(\dot{\boldsymbol{v}}_i,\boldsymbol{\psi})$ can be expressed as
\begin{equation*}
s(\dot{\boldsymbol{v}}_i,\boldsymbol{\psi}) = \frac{1}{2} \sum_t \left(\dot{y}_{it} - \dot{\boldsymbol{z}}_i'\boldsymbol{\pi}_{t}(\boldsymbol{\psi})\right)^2
\end{equation*}
where the entries in the $m \times 1$ vector $ \boldsymbol{\pi}_{t}(\boldsymbol{\psi}) $ are nonlinear functions of the entries in $\boldsymbol{\psi}$. If the variables in $\dot{\boldsymbol{z}}_i$ are not perfectly collinear, it follows from well-known least squares theory that $\boldsymbol{\pi}_{0t} := \boldsymbol{\pi}_{t}(\boldsymbol{\psi}_0)$ is identified for each $t$.  Therefore, $\boldsymbol{\psi}_0$ is identified if it can be recovered from the $\boldsymbol{\pi}_{0t}$s.  More formally, given that the variables in $\dot{\boldsymbol{z}}_i$ are not perfectly collinear, the necessary and sufficient condition for identification of $\boldsymbol{\psi}_0$ is that, for each $\boldsymbol{\psi} \in \boldsymbol{\Psi}$ such that $\boldsymbol{\psi} \ne \boldsymbol{\psi}_0$ there is at least one $t \in \{1,\ldots,T\}$ such that $\boldsymbol{\pi}_t(\boldsymbol{\psi}) \ne \boldsymbol{\pi}_t(\boldsymbol{\psi}_0)$. That this condition is necessary and sufficient for identification is established  as part of the proof of Theorem \ref{consistency_thm}. 

Example 2 illustrates identification.  It shows that because the $\lambda_{0t}$s and $\theta_{0j}$s are interacted, identification of the $\lambda_{0t}$s obviously requires some of the  $\theta_{0j}$s to be nonzero.  Of course, the latter will be true whenever some of the regressors are correlated with the effects in $\boldsymbol{\widetilde{\eta}}_i$.

\subsubsection*{Example 2. }
Suppose that $K=1$, $q=1$, and $\boldsymbol{z}_i=(x_{i1},\ldots, x_{iT})$. In this case, $\boldsymbol{\Lambda}_0' = (1, \lambda_{02},\ldots, \lambda_{0T})$. For $t=1,\ldots,m=T$, let $\theta_{0t}$ denote the $t$th component of $\boldsymbol{\theta}_{0}$, i.e., the coefficient on $x_{it}$ in the projection of $\eta_i$ on $\boldsymbol{z}_i$.  Then, after demeaning the variables, we have 
\begin{equation*}
\begin{split}
\dot{y}_{it} & = \beta_0 \dot{x}_{it} + \lambda_{0t} \dot{\boldsymbol{z}}_i'\boldsymbol{\theta}_{0} + \dot{u}_{it} \\
& = \beta_0 \dot{x}_{it} + \lambda_{0t}\theta_{01}\dot{x}_{i1} + \cdots + \lambda_{0t}\theta_{0t}\dot{x}_{it} + \cdots + \lambda_{0t}\theta_{0T}\dot{x}_{iT} + \dot{u}_{it} \\
& = \lambda_{0t}\theta_{01}\dot{x}_{i1} + \cdots + \left(\beta_0 + \lambda_{0t}\theta_{0t}\right)\dot{x}_{it} + \cdots + \lambda_{0t}\theta_{0T}\dot{x}_{iT} + \dot{u}_{it} \\
& = \dot{\boldsymbol{z}}_{i}'\boldsymbol{\pi}_{0t} + \dot{u}_{it},
\end{split}
\end{equation*}
where $\boldsymbol{\pi}_{01}'=(\beta_0 + \theta_{01},\theta_{02},\ldots,\theta_{0T})$ and $\boldsymbol{\pi}_{0t}'=(\lambda_{0t}\theta_{01},\ldots, \beta_0+\lambda_{0t}\theta_{0t},\ldots,\lambda_{0t}\theta_{0T})$ for $t =2,\ldots,T$. Given $\boldsymbol{\pi}_{01}$ is identified (in the absence of collinearity in $\boldsymbol{z}_i$), the parameters $\theta_{02},\ldots, \theta_{0T}$ are identified. Then, if at least two of these parameters are nonzero, we can identify $\lambda_{0t}$ for $t=2,\ldots,T$ from $\boldsymbol{\pi}_{0t}$. To see this, suppose $\theta_{02} \ne 0$, then, because $\theta_{02} $ is identified and the products $\lambda_{0t}\theta_{02}$ are identified for $t \ne2$, we can recover $\lambda_{0t}$ for $t \ne 2$. Only $\lambda_{02}$, $\theta_{01}$, and $\beta_{0}$ remain unidentified.  But they can be identified if one more $\theta_{0t}$ is nonzero.  Suppose, for example, that $\theta_{03} \ne 0$. Because both $\theta_{03}$ and the product $\lambda_{02}\theta_{03}$ are identified, we can recover $\lambda_{02}$.  The parameter $\beta_0$ can then be identified from the $t$th component of $\boldsymbol{\pi}_{0t}$ for any $t\geq2$. Once $\beta_0$ is known, we can recover $\theta_{01}$ from the first component of $\boldsymbol{\pi}_{01}$. The condition that at least two of the parameters, $\theta_{02},\ldots, \theta_{0T}$, are nonzero is a necessary and sufficient condition for identification in this case.\\

A necessary condition for identification consists of simply counting parameters. To see this, note that each $\boldsymbol{\pi}_{0t}$ contains $m$ parameters, and there are $T$ $\boldsymbol{\pi}_{0t}$ vectors, for a total of $mT$ identifiable parameters in the $\boldsymbol{\pi}_{0t}$s.  On the other hand, there are $K+q(m+T-q)$ parameters in $\boldsymbol{\psi}_0$.  Clearly, we cannot recover more entries in $\boldsymbol{\psi}_{0}$ than the total number of entries in the $\boldsymbol{\pi}_{0t}$ vectors.  In other words, we must have $mT \ge K+q(m+T-q)$, or
\begin{equation} \label{identification}
	(T-q)(m-q) \ge K.
\end{equation}

This necessary condition can be used to derive non-identification results. To illustrate this point, consider Example 1 again.

\subsubsection*{Example 1 revisited.}
In the ``treatment effect'' model with $T=2$ and $z_i=x_{i2}$ is scalar, we can immediately conclude from (\ref{identification}) that the model is not identified.  This is because $m=1$, and thus $ m - q = 0 $.  Indeed, for this model, after demeaning we have $\dot{y}_{it} = \pi_{0t}\dot{x}_{i2} + \dot{u}_{it}$, for $t=1,2$, with $\pi_{01}=\lambda_{01}\theta_0$ and $\pi_{02}=\beta_0+\lambda_{02}\theta_0$. Given the normalization $\lambda_{01} = 1$, we can recover $\theta_0$ from $\pi_{01}=\theta_0$. However, even if $\theta_0$ and $\pi_{02}$ are known, we cannot recover both $\beta_0$ and $\lambda_{02}$.  Furthermore, extending the number of time periods to $T>2$ does not improve the situation if treatment assignment does not change in subsequent periods. That is, if $x_{it}=0$ for $t<t_0$ and $x_{it}=x_{it_0}$ for all $t\geq t_0$ then $z_i=x_{it_0}$ is still a scalar ($m=1$) and the model is not identified. 

There are two ways to achieve identification. One possibility is staggered treatment adoption.  If $T=3$ and some individuals  receive treatment in period 2 while others receive it in period 3, then $\dot{y}_{it} = \boldsymbol{z}_i'\boldsymbol{\pi}_{0t} + \dot{u}_{it}$, $t=1,2,3$, with $\dot{\boldsymbol{z}}_i'=(\dot{x}_{i2},\dot{x}_{i3})$, $\boldsymbol{\pi}_{01}' = (\lambda_{01}\theta_{01},\lambda_{01}\theta_{02})$, $\boldsymbol{\pi}_{02}' = (\beta_{0}+\lambda_{02}\theta_{01},\lambda_{02}\theta_{02})$,  and $\boldsymbol{\pi}_{03}' = (\lambda_{03}\theta_{01},\beta_0 + \lambda_{03}\theta_{02})$.  There are a total of six entries in the $\boldsymbol{\pi}_{0t}$s, but, given the normalization $\lambda_{01}=1$, there are only five parameters $\theta_{01}$,$\theta_{02}$, $\lambda_{02}$, $\lambda_{05}$, and $\beta_0$ that need to be recovered from the $\boldsymbol{\pi}_{0t}$s.   To identify the latter parameters, it suffices that $\theta_{01}$ and $\theta_{02}$ are both nonzero. Note that, given $\lambda_{01}=1$, these parameters can be recovered from $\boldsymbol{\pi}_{01}' = (\theta_{01},\theta_{02})$; $\lambda_{02}$ can then be recovered from the second entry in $\boldsymbol{\pi}_{02}$; $\lambda_{03}$ can then be recovered from the first entry in $\boldsymbol{\pi}_{03}$; and, finally, $\beta_0$ can be recovered from either the first entry of $\boldsymbol{\pi}_{02}$, of the second entry of $\boldsymbol{\pi}_{03}$.

Another solution is when there is another regressor that is time-varying. With sufficient variation in this regressor we can identify $\boldsymbol{\Lambda}_{0}$, as in Example 2. The component of $\boldsymbol{\pi}_{0t}$ corresponding to the ``treatment'' variable will be $\lambda_{0t}\theta_{01}$ for $t<t_0$ and $\beta_0+\lambda_{0t}\theta_{01}$ for $t\geq t_0$.  Then, given the normalization $\lambda_{01}=1$, $\theta_{01}$ is identified from $t=1$, and $\beta_0$ is identified from any $t\geq t_0$.

\subsection{Large sample properties  \label{estimation}}

\fussy

Theorem \ref{consistency_thm} provides conditions for  consistent estimation of $  \boldsymbol{\psi}_0 $ via nonlinear least squares.

\begin{theorem} \label{consistency_thm}
	Assume $ \boldsymbol{\Phi}_{01} $ is nonsingular and 
	\begin{itemize} 
		\item [A1:] $ \boldsymbol{v}_i :=(\boldsymbol{z}_i', \boldsymbol{a}_i', \boldsymbol{\epsilon}_i')' $ $(i=1,\ldots,n)$ are independently and identically distributed (i.i.d.);
		\item [A2:] the entries in $\boldsymbol{v}_i$ have finite second moments;
		\item [A3:] $ E(\boldsymbol{z}_1 \boldsymbol{\epsilon}_{1}^{\prime}) = \boldsymbol{0}$;
		\item [A4:] $ \text{Var}(\boldsymbol{z}_1)  $ is positive definite; and
		\item [A5:] for each $\boldsymbol{\psi} \in \boldsymbol{\Psi}$ such that $\boldsymbol{\psi} \ne \boldsymbol{\psi}_0$ there is at least one $t \in \{1,\ldots,T\}$ such that $\boldsymbol{\pi}_t(\boldsymbol{\psi}) \ne \boldsymbol{\pi}_t(\boldsymbol{\psi}_0)$.
	\end{itemize}
	Then there is a measurable minimizer, $ \widehat{\boldsymbol{\psi}} $, of $ S_{n}(\cdot)  $ in $\overline{\boldsymbol{\Psi}}$, a compact subset of $\boldsymbol{\Psi}$ with $\boldsymbol{\psi}_0 $ in its interior, such that
	\begin{equation*}
		\widehat{\boldsymbol{\psi}} \rightarrow_p \boldsymbol{\psi}_0 \qquad (n \rightarrow \infty, \;  T \text{ fixed}).
	\end{equation*}
\end{theorem}

Proofs are provided in Appendix A.

As is true for the consistency of the TWFE estimator, the thoerem imposes no restrictions on the behavior of the idiosyncratic errors --- the $ \epsilon_{it} $s.  For a given individual, they can be conditionally heteroskedastic, time-series heteroskedastic, or correlated in an arbitrary fashion.

Under only slightly stronger assumptions, asymptotic normality of the NLS estimator can be verified.  Theorem \ref{normality_thm}, which establishes asymptotic normality, relies on a few additional definitions.  Specifically, let $ \boldsymbol{X}_{1i} $ be the first $ q $ rows of $ \boldsymbol{X}_i $ and let $ \boldsymbol{X}_{2i} $ be the last $ (T-q) $ rows.  Also, let 
\begin{equation}\label{Rmatrix}
\boldsymbol{R}_i := \left(  
\begin{array}{ccc}
 \boldsymbol{X}_{1i} & \boldsymbol{Z}_i  &  \boldsymbol{0} \\
 \boldsymbol{X}_{2i} & \boldsymbol{\Lambda}_{02} \boldsymbol{Z}_i &  \boldsymbol{\theta}_0^{\prime}\boldsymbol{Z}_i^{\prime} \otimes \boldsymbol{I}_{T-q}   
\end{array}
\right) \quad \text{and} \quad \boldsymbol{R}_i^{\ast} := \boldsymbol{R}_i - E\left( \boldsymbol{R}_1\right).
\end{equation}
Finally, define $ \boldsymbol{H}_0 = E \left( \boldsymbol{R}_{1}^{\ast \prime} \boldsymbol{R}_{1}^{\ast}\right)  $.

We can now state Theorem \ref{normality_thm}. 

\begin{theorem} \label{normality_thm}
	Assume $ \boldsymbol{\Phi}_{01} $ is nonsingular, A1 and A3--A5 are satisfied, and 
	\begin{itemize} 
		\item [A2$^{\ast}$:] the entries in $\boldsymbol{v}_i$ have finite fourth moments.
	\end{itemize}
	Then $ \boldsymbol{A}_0 = E \left(  \boldsymbol{R}_1^{\ast \prime}\boldsymbol{u}_1\boldsymbol{u}_1^{\prime}\boldsymbol{R}_1^{\ast} \right)  $ has finite entries, $\boldsymbol{H}_0$ is positive definite, and 
	\begin{equation*}
		\sqrt{n}\left( \widehat{\boldsymbol{\psi}} - \boldsymbol{\psi}_0 \right) \rightarrow_d N\left(\boldsymbol{0},\boldsymbol{H}_0^{-1}\boldsymbol{A}_0\boldsymbol{H}_0^{-1}\right) \qquad (n \rightarrow \infty, \;  T \text{ fixed}).
	\end{equation*}
\end{theorem}

Under the conditions of Theorem \ref{normality_thm}, the global identification condition in Assumption A5 implies the matrix $\boldsymbol{H}_0$ is positive definite.  On the other hand, due to the nonlinearity of the model, $\boldsymbol{H}_0$ being positive definite does not ensure global identification.  However, if the matrix $ \boldsymbol{H}_0$ is positive definite, the model is \textit{locally} identified. This rank condition is similar to Assumption BA.4 in \cite{Ahn2013}. Such rank conditions are useful for developing a test of under-identification but do not provide insights into model identification that cannot alternatively be gleaned from the global identification condition  or from the simple necessary condition in (\ref{identification}).

\section{Testing for the consistency of TWFE \label{test}}

As already noted in Section \ref{model}, the presence of interactive effects, by itself,  does not necessarily imply the TWFE estimator is inconsistent.  Even if  interactive effects are present, the TWFE estimator can still be consistent should all of the covariances in $\text{Cov}(\widetilde{\boldsymbol{\eta}}_i,\boldsymbol{x}_{it}) $ be time invariant---in other words, if  $\boldsymbol{\Phi}_{0t}^{x} = \boldsymbol{\Phi}_{0}^{x}$ for all $t$ (see Eq.s (\ref{x_ie}) and (\ref{bias})). On the other hand, an important motivation for considering the possibility of interactive effects is that they can render the TWFE estimator inconsistent when some of the covariances in $\text{Cov}(\widetilde{\boldsymbol{\eta}}_i,\boldsymbol{x}_{it}) $ vary across time.  

Theorem \ref{dist_4_test_stat} provides a joint distribution result that can be used to derive a statistic for testing for the consistency of the TWFE estimator.  The TWFE estimator is consistent if there are no interactive effects---that is, if $q=1$ and $\boldsymbol{\Lambda}_0 = \boldsymbol{\iota}_T$.  Moreover, as already noted, it is also consistent if $\boldsymbol{\Phi}_{0t}^{x} = \boldsymbol{\Phi}_{0}^{x}$ for all $t$. Therefore, under the null hypothesis that (A) $\boldsymbol{\Lambda}_0 = \boldsymbol{\iota}_T$ is true, or (B) $\boldsymbol{\Phi}_{0t}^{x} = \boldsymbol{\Phi}_{0}^{x}$ for all $t$ is true, or both (A) and (B) are true, the PIE estimator and the TWFE estimator are both consistent, whereas only the PIE estimator is consistent if the null hypothesis is false. This fact implies the consistency of the TWFE can be tested using contrasts between the PIE and TWFE estimators, provided we know the joint distribution of these estimators under the null hypothesis. Theorem \ref{dist_4_test_stat} gives the joint distribution for our new estimator $\widehat{\boldsymbol{\psi}}$ and the TWFE estimator $\widehat{\boldsymbol{\beta}}_{FE}$, assuming the null hypothesis is true.

 \begin{theorem} \label{dist_4_test_stat}
 Let $q = 1$. Then $\boldsymbol{\Lambda}_0' = (1,\,\boldsymbol{\lambda}_0')$.  Assume  A1, A2$^{\ast}$, and A3--A5 are satisfied.  If (A) $\boldsymbol{\Lambda}_0 = \boldsymbol{\iota}_T$ is true, or (B) $\boldsymbol{\Phi}_{0t}^{x} = \boldsymbol{\Phi}_{0}^{x}$ for all $t$ is true, or both (A) and (B) are true, then
 \begin{equation*}
\sqrt{n}
\left(
\begin{array}{c}
  \widehat{\boldsymbol{\psi}} - \boldsymbol{\psi}_0   \\
  \widehat{\boldsymbol{\beta}}_{FE} - \boldsymbol{\beta}_0
 \end{array}
\right)
\rightarrow_d N\left(\,\boldsymbol{0},\,(\boldsymbol{H}_0^+)^{-1}\boldsymbol{A}_0^+(\boldsymbol{H}_0^+)^{-1}\,\right) \quad   (n \rightarrow \infty, \;  T \text{ fixed}),
\end{equation*}
where $ \boldsymbol{H}_0^+ = E \left( \boldsymbol{R}_1^{+ \prime} \boldsymbol{R}_1^+\right)  $, $\boldsymbol{R}_i^+ := \text{diag}(\boldsymbol{R}_i^{\ast}, \boldsymbol{Q}\boldsymbol{X}_i^{\ast})$,
$ \boldsymbol{A}_0^+ := E \left(  \boldsymbol{R}_1^{+ \prime}\boldsymbol{u}_1^+\boldsymbol{u}_1^{+ \prime}\boldsymbol{R}_1^{+}\right)  $, and $\boldsymbol{u}_i^{+ \prime} := (\boldsymbol{u}_i', (\boldsymbol{y}_i^{\ast} - \boldsymbol{X}_i^{\ast}\boldsymbol{\beta}_0)')$
\end{theorem}

A statistic for testing the consistency of the TWFE estimator can be derived by exploiting the conclusion of Theorem \ref{dist_4_test_stat}. To construct the statistic, let $\boldsymbol{C} := ( \begin{array}{cccc}
		\boldsymbol{I}_{K} & \boldsymbol{0} & -\boldsymbol{I}_{K}\end{array})$
be such that it picks out $ \widehat{\boldsymbol{\beta}}  - \widehat{\boldsymbol{\beta}}_{FE}  $ when it premultiplies $ (\widehat{\boldsymbol{\psi}}^{\prime},\widehat{\boldsymbol{\beta}}_{FE}^{\prime})^{\prime}  $.  Then, Theorem \ref{dist_4_test_stat} implies that if the statement (A) $\boldsymbol{\Lambda}_0 = \boldsymbol{\iota}_T$ is true, or the statement (B) $\boldsymbol{\Phi}_{0t}^{x} = \boldsymbol{\Phi}_{0}^{x}$ for all $t$  is true, or both statements (A) and (B) are true, the random variable 
\begin{equation} \label{Wald}
	\vartheta_{n} := n\left( \widehat{\boldsymbol{\beta}} - \widehat{\boldsymbol{\beta}}_{FE} \right)^{\prime} \left( \boldsymbol{C} (\boldsymbol{H}_0^+)^{-1}\boldsymbol{A}_0^+ (\boldsymbol{H}_0^+)^{-1}\boldsymbol{C}^{\prime}\right) ^{-1}\left( \widehat{\boldsymbol{\beta}} - \widehat{\boldsymbol{\beta}}_{FE}\right),
\end{equation}
 has a chi-square distribution with $K$ degrees of freedom, as $n \rightarrow \infty$.  On the other hand, given the TWFE estimator is inconsistent under the alternative that (A) and (B) are both false, while the PIE estimator remains consistent , $\vartheta_{n}$ will tend to be too large when the null hypothesis is false.
 
 Of course, $\vartheta_{n}$ is not a test statistic until $ (\boldsymbol{H}_0^+)^{-1}\boldsymbol{A}_0^+ (\boldsymbol{H}_0^+)^{-1}$ in (\ref{Wald}) is replaced with a consistent estimator.  To that end,  note that $\boldsymbol{H}_0^+$ is consistently estimated with $ \widehat{\boldsymbol{H}}^+ := (1/n)\sum_i\widehat{\boldsymbol{R}}_i^{+ \prime}\widehat{\boldsymbol{R}}_i^+$, where $\widehat{\boldsymbol{R}}_i^+ := \text{diag}(\widehat{\boldsymbol{R}}_i,\, \boldsymbol{Q}\dot{\boldsymbol{X}}_i)$, and
 \begin{equation} \label{Rmatrix_est}
\widehat{\boldsymbol{R}}_i := \left(  
\begin{array}{ccc}
 \dot{\boldsymbol{X}}_{1i} & \dot{\boldsymbol{Z}}_i  &  \boldsymbol{0} \\
 \dot{\boldsymbol{X}}_{2i} & \widehat{\boldsymbol{\Lambda}}_{2} \dot{\boldsymbol{Z}}_i &  
\widehat{\boldsymbol{\theta}}^{\prime}\dot{\boldsymbol{Z}}_i^{\prime} \otimes \boldsymbol{I}_{T-q}   
\end{array}
\right)
\end{equation}
is an estimator of $\boldsymbol{R}_i^{\ast}$ (see (\ref{Rmatrix})).  In (\ref{Rmatrix_est}), $\widehat{\boldsymbol{\Lambda}}_{2}$ and $\widehat{\boldsymbol{\theta}}$ are the NLS estimators of $\boldsymbol{\Lambda}_{02}$ and $\boldsymbol{\theta}_0$. Moreover, $\boldsymbol{A}_0^+$ can be estimated with $\widehat{\boldsymbol{A}}^+ := (1/n)\sum_i \widehat{\boldsymbol{R}}_i^{+ \prime}\widehat{\dot{\boldsymbol{u}}}_i^+\widehat{\dot{\boldsymbol{u}}}_i^{+ \prime}\widehat{\boldsymbol{R}}_i^+$, where  $\widehat{\dot{\boldsymbol{u}}}_i^{+ \prime} := (\widehat{\dot{\boldsymbol{u}}}_i^{ \prime}, (\dot{\boldsymbol{y}}_i - \dot{\boldsymbol{X}}_i\boldsymbol{b})')$, $\widehat{\dot{\boldsymbol{u}}}_i := \dot{\boldsymbol{y}}_i - \dot{\boldsymbol{X}}_i\widehat{\boldsymbol{\beta}} -  \widehat{\boldsymbol{\Lambda}}\dot{\boldsymbol{Z}}_i\widehat{\boldsymbol{\theta}}$, and $\boldsymbol{b}$ is a consistent estimator of $\boldsymbol{\beta}_0$. The latter can be either the PIE estimator, $\widehat{\boldsymbol{\beta}}$, or the TWFE estimator, $\widehat{\boldsymbol{\beta}}_{FE}$, because both are consistent when the null hypothesis is true.  Replacing $(\boldsymbol{H}_0^+)^{-1}\boldsymbol{A}_0^+ (\boldsymbol{H}_0^+)^{-1}$ with $ (\widehat{\boldsymbol{H}}^+)^{-1}\widehat{\boldsymbol{A}}^+  (\widehat{\boldsymbol{H}}^+)^{-1}$ in (\ref{Wald}) yields the relevant statistic for testing for the consistency of the TWFE estimator.
 
 \fussy

\section{Computation \label{computation}}
The NLS estimate $ \widehat{\boldsymbol{\psi}}  = ( \widehat{\boldsymbol{\beta}}', \,
\widehat{\boldsymbol{\theta}}', \, \widehat{\boldsymbol{\lambda}}') '$  can be found by minimizing the objective function $ S_{n}(\boldsymbol{\psi}) $.  Although that computational strategy is possible, minimizing $ S_{n}(\boldsymbol{\psi}) $ directly with respect to $ \boldsymbol{\psi} $ is a high-dimensional, nonlinear optimization problem.  This section shows how $ \widehat{\boldsymbol{\beta}} $ and $ \widehat{\boldsymbol{\lambda}} $ can be calculated by solving a simpler computational problem.

The simpler computational solution relies on the conclusion of Theorem \ref{computation_thm} which provides an alternative characterization of $ \widehat{\boldsymbol{\psi}}$.  In order to state that theorem, a few additional definitions are needed.  Specifically,  let $\dot{\boldsymbol{y}}=(\dot{\boldsymbol{y}}_1',\ldots,\dot{\boldsymbol{y}}_  n')'$, $\dot{\boldsymbol{X}}:=(\dot{\boldsymbol{X}}_1',\ldots,\dot{\boldsymbol{X}}_n')'$, and $\dot{\boldsymbol{z}}:=(\dot{\boldsymbol{z}}_1,\ldots,\dot{\boldsymbol{z}}_n)'$.  Also, let $\boldsymbol{P}_{\dot{\boldsymbol{X}}}$, $\boldsymbol{P}_{\boldsymbol{\Lambda} }$, and $\boldsymbol{P}_{\dot{\boldsymbol{z}} }$ denote the projection matrices $\boldsymbol{P}_{\dot{\boldsymbol{X}}} := \dot{\boldsymbol{X}}(\dot{\boldsymbol{X}}'\dot{\boldsymbol{X}} )^{-1}\dot{\boldsymbol{X}}'  $, $ \boldsymbol{P}_{\boldsymbol{\Lambda} } := \boldsymbol{\Lambda}\left(\boldsymbol{\Lambda}'\boldsymbol{\Lambda}\right)^{-1}\boldsymbol{\Lambda}' $, and $ \boldsymbol{P}_{\dot{\boldsymbol{z}} }:= \dot{\boldsymbol{z}}(\dot{\boldsymbol{z}}'\dot{\boldsymbol{z}})^{-1}\dot{\boldsymbol{z}}' $.  Finally, set $ \boldsymbol{M} := \boldsymbol{I}_{nT} - \boldsymbol{P}_{\dot{\boldsymbol{X}}}$ and $
\boldsymbol{M}(\boldsymbol{\lambda}):=\boldsymbol{I}_{nT}-\left(\boldsymbol{P}_{\dot{\boldsymbol{z}} } \otimes \boldsymbol{P}_{\boldsymbol{\Lambda}} \right)
$.

\begin{theorem} \label{computation_thm} 
Let  $\boldsymbol{\alpha} := (\boldsymbol{\beta}' ,\boldsymbol{\lambda}')'$ and define $\widetilde{\boldsymbol{\alpha}} = (\widetilde{\boldsymbol{\beta}}' ,\widetilde{\boldsymbol{\lambda}}')'$ as 
\begin{equation} \label{FWL_obj_fcn}
	\widetilde{\boldsymbol{\alpha}} := \arg\min_{\boldsymbol{\alpha}}\left(\dot{\boldsymbol{y}} - \dot{\boldsymbol{X}}\boldsymbol{\beta} \right)'\boldsymbol{M}(\boldsymbol{\lambda})\left(\dot{\boldsymbol{y}} - \dot{\boldsymbol{X}}\boldsymbol{\beta} \right).
\end{equation}
Then $\widetilde{\boldsymbol{\beta}}$ must be 
\begin{equation} \label{PIE_soln}
	\widetilde{\boldsymbol{\beta}}  =  \left( \sum_i \dot{\boldsymbol{X}}_i'\boldsymbol{Q}(\widetilde{\boldsymbol{\lambda}}) \dot{\boldsymbol{X}}_i\right) ^{-1} \sum_i \dot{\boldsymbol{X}}_i'\boldsymbol{Q}(\widetilde{\boldsymbol{\lambda}}) \dot{\boldsymbol{y}}_i,
\end{equation}
with $ \boldsymbol{Q}(\boldsymbol{\lambda}) := \boldsymbol{I}_T - \boldsymbol{P}_{\boldsymbol{\Lambda} }  $. Moreover, let the entries in the $m \times q$ matrix $\widehat{\boldsymbol{\Theta}}^o(\boldsymbol{\lambda})$ be such that 
\begin{equation} \label{FWL_theta}
	  \text{vec}(\widehat{\boldsymbol{\Theta}}^o(\boldsymbol{\lambda})') =  \left( (\dot{\boldsymbol{z}} \otimes \boldsymbol{\Lambda})'\boldsymbol{M}(\dot{\boldsymbol{z}} \otimes \boldsymbol{\Lambda})\right)^{-1}(\dot{\boldsymbol{z}} \otimes \boldsymbol{\Lambda})'\boldsymbol{M}\dot{\boldsymbol{y}}.
\end{equation}
Finally, set $\widetilde{\boldsymbol{\theta}} := \text{vec}(\widehat{\boldsymbol{\Theta}}^o(\widetilde{\boldsymbol{\lambda}}))$ and 
$ \widetilde{\boldsymbol{\psi}} := (\widetilde{\boldsymbol{\beta}}', \widetilde{\boldsymbol{\theta}}',\widetilde{\boldsymbol{\lambda}}')'$.
Then
\begin{equation} \label{min_S_n}
	S_{n}(\widetilde{\boldsymbol{\psi}}) \le S_{n}(\boldsymbol{\psi}) \quad \text{for all }\boldsymbol{\psi},
\end{equation}
and therefore $\widehat{\boldsymbol{\psi}}=\widetilde{\boldsymbol{\psi}}$.
\end{theorem}

Because $\widetilde{\boldsymbol{\psi}}$ minimizes $S_{n}(\cdot)$, and therefore must be the NLS estimate (i.e. $\widehat{\boldsymbol{\psi}}=\widetilde{\boldsymbol{\psi}}$), Theorem~\ref{computation_thm} shows that the NLS estimates $\widehat{\boldsymbol{\beta}}$ and $\widehat{\boldsymbol{\lambda}}$ can be found by solving  (\ref{FWL_obj_fcn}).  The parameters in  $\boldsymbol{\theta}$ do not appear in the minimization problem in (\ref{FWL_obj_fcn}).  On the other hand, Theorem~\ref{computation_thm} also shows that if we want an estimate of $\boldsymbol{\theta}_0$, it can be recovered from Eq. (\ref{FWL_theta}) after the estimate $\widehat{\boldsymbol{\lambda}}$ is calculated.  Moreover, as stated in the theorem, the solution to  (\ref{FWL_obj_fcn}) must be such that Eq. (\ref{PIE_soln}) is satisfied. 

The preceding observations beg the question, how can the objective function in (\ref{FWL_obj_fcn}) be minimized?  To see how that function can be minimized, we show that for a given  $\boldsymbol{\beta}$ there is a computationally tractable solution for $\boldsymbol{\lambda}$, justifying an iterative procedure similar to the procedure for the large-$T$ IE model originally proposed by \cite{Kiefer1980} (see also \cite{Bai2009}). 

\sloppy
Specifically,  let $ \boldsymbol{\lambda}^{(i)} $  denote the fit of $ \boldsymbol{\lambda} $ on the $i$th iteration. Then $(\dot{\boldsymbol{y}} - \dot{\boldsymbol{X}}\boldsymbol{\beta})'\boldsymbol{M}(\boldsymbol{\lambda}^{(i)})(\dot{\boldsymbol{y}} - \dot{\boldsymbol{X}}\boldsymbol{\beta})$ is minimized with respect to $\boldsymbol{\beta}$ by $ \boldsymbol{\beta}^{(i)} = \widehat{\boldsymbol{\beta}}(\boldsymbol{\lambda}^{(i)}) $, where
\begin{equation} \label{updating_PIE}
	\widehat{\boldsymbol{\beta}}(\boldsymbol{\lambda})  =  \left( \sum_i \dot{\boldsymbol{X}}_i'\boldsymbol{Q}(\boldsymbol{\lambda}) \dot{\boldsymbol{X}}_i\right) ^{-1} \sum_i \dot{\boldsymbol{X}}_i'\boldsymbol{Q}(\boldsymbol{\lambda}) \dot{\boldsymbol{y}}_i.
\end{equation}
Once $\boldsymbol{\beta}^{(i)}$ is calculated, the sum  $ (\dot{\boldsymbol{y}} - \dot{\boldsymbol{X}}\boldsymbol{\beta}^{(i)} )'\boldsymbol{M}(\boldsymbol{\lambda})(\dot{\boldsymbol{y}} - \dot{\boldsymbol{X}}\boldsymbol{\beta}^{(i)} ) $  can be minimized with respect to $ \boldsymbol{\lambda} $ to get a new fit for $ \boldsymbol{\lambda} $. 
\fussy

To see how $ (\dot{\boldsymbol{y}} - \dot{\boldsymbol{X}}\boldsymbol{\beta} )'\boldsymbol{M}(\boldsymbol{\lambda})(\dot{\boldsymbol{y}} - \dot{\boldsymbol{X}}\boldsymbol{\beta}) $ is minimized with respect to $ \boldsymbol{\lambda} $,  let $\dot{\boldsymbol{e}}(\boldsymbol{\beta}) :=\dot{\boldsymbol{y}}-\dot{\boldsymbol{X}}\boldsymbol{\beta}$ and let $\boldsymbol{E}(\boldsymbol{\beta})$ be the $n\times T$ matrix such that $\text{vec}(\boldsymbol{E}(\boldsymbol{\beta})')=\dot{\boldsymbol{e}}(\boldsymbol{\beta})$. We show in Appendix B that the matrix $\boldsymbol{\Lambda}$ that minimizes $ (\dot{\boldsymbol{y}} - \dot{\boldsymbol{X}}\boldsymbol{\beta})'\boldsymbol{M}(\boldsymbol{\lambda})(\dot{\boldsymbol{y}} - \dot{\boldsymbol{X}}\boldsymbol{\beta}) $ must solve 
\begin{equation}\label{step2}
\max_{\boldsymbol{\Lambda}^{\ast}} \text{tr}\left( \boldsymbol{\Lambda} ^{\ast \prime}\boldsymbol{\Sigma}(\boldsymbol{\beta})\boldsymbol{\Lambda} ^{\ast}\right) \quad \text{subject to} \quad \boldsymbol{\Lambda} ^{\ast }=\boldsymbol{\Lambda} (\boldsymbol{\Lambda} '\boldsymbol{\Lambda} )^{-1/2}
\end{equation} 
where $\boldsymbol{\Sigma}(\boldsymbol{\beta}):=\boldsymbol{E}(\boldsymbol{\beta})'\boldsymbol{P}_{\dot{\boldsymbol{z}}}\boldsymbol{E}(\boldsymbol{\beta})$. This is solved by setting $\boldsymbol{\Lambda} ^{\ast }$ equal to the $q$ eigenvectors associated with the $q$ largest eigenvalues of $\boldsymbol{\Sigma}(\boldsymbol{\beta})$.\footnote{Noting that $\boldsymbol{\Lambda} ^{\ast \prime} \boldsymbol{\Lambda} ^{\ast }=I_q$, this is a well-known result; see, for example, \cite{Anderson1984}.} After  the fit of $\boldsymbol{\Lambda} ^{\ast }$ is calculated, the appropriate fit of $\boldsymbol{\Lambda}$ is obtained by imposing the normalization that the first $q$ rows are equal to the identity matrix.

The algorithm is summarized  as follows. 
\begin{itemize}
\item[1. ] Remove time fixed effects by subtracting time-specific cross-sectional means from each variable. Set $i=0$.
\item[2. ] Set $\boldsymbol{\beta}=\boldsymbol{\beta}^{(i)}$ and use (\ref{step2}) to solve for $\boldsymbol{\lambda}^{(i+1)}$.
\item[3. ] Set $\boldsymbol{\lambda}=\boldsymbol{\lambda}^{(i+1)}$ and use (\ref{updating_PIE}) to solve for $\boldsymbol{\beta}^{(i+1)}$.
\item[4. ] If $||\boldsymbol{\beta}^{(i+1)}-\boldsymbol{\beta}^{(i)}||<\epsilon$, stop. Otherwise, return to Step 2 with $i \leftarrow i+1$. 
\end{itemize}

The foregoing reveals that the small-$ T $ PIE estimator of $ \boldsymbol{\beta}_0 $ can be expressed as
\begin{equation} \label{PIE}
	\widehat{\boldsymbol{\beta}}  =  \left( \sum_i \dot{\boldsymbol{X}}_i'\boldsymbol{Q}(\widehat{\boldsymbol{\lambda}}) \dot{\boldsymbol{X}}_i\right) ^{-1} \sum_i \dot{\boldsymbol{X}}_i'\boldsymbol{Q}(\widehat{\boldsymbol{\lambda}}) \dot{\boldsymbol{y}}_i.
\end{equation}
The fixed effect estimator for the IE model can also be expressed as in equation (\ref{PIE}) \citep[see, e.g.,][]{Lee1991, Ahn2001, Bai2009}. But it is computed by iterating between (\ref{updating_PIE}) and a modification of (\ref{step2}), which consists of using  $\boldsymbol{E}(\boldsymbol{\beta})'\boldsymbol{E}(\boldsymbol{\beta})$ instead of $\boldsymbol{E}(\boldsymbol{\beta})'\boldsymbol{P}_{\dot{\boldsymbol{z}}}\boldsymbol{E}(\boldsymbol{\beta})$. As noted by \cite{Bai2009}, convergence of such an iterative procedure to a local optimum was shown by \cite{Sargan1964}.\footnote{We also verify convergence of the iterative algorithm in our simulations.}

An advantage of  the small-$ T $ PIE estimator of $ \boldsymbol{\beta}_0 $ is that it is robust with respect to heteroskedasticity and serial correlation in the errors, even when $ T $ is small.  On the other hand, when $ T $ is fixed, the consistency of the large-$ T $ IE estimator of $ \boldsymbol{\beta}_0 $ depends on the errors being conditionally homoskedastic and uncorrelated (see, e.g., Ahn et al., 2001). Unlike in the AE model, the incidental parameter $ \boldsymbol{\lambda}_0 $ must be estimated consistently. The large-$T$ IE estimator relies on strong serial dependence in the error structure produced by the presence of the common factors to identify these factor loadings, and as a result requires the residual dependence to be weak. In contrast, the PIE estimator relies on dependence between the errors and the regressors to identify the factor loadings.

As already noted, calculating the NLS estimates of $\boldsymbol{\beta}_0$ and $\boldsymbol{\lambda}_0$ with the iterative procedure described in this section does not rely on computing an estimate of $ \boldsymbol{\theta}_0$.  This is because the conclusion of Theorem \ref{computation_thm} follows from applying the FWL theorem, and, as is well-known, the FWL theorem shows how estimates of a subset of parameters in a parameter vector can be calculated without calculating estimates of the remaining parameters in the vector.  
   
On the other hand,  testing for the consistency of the TWFE estimator, as described in Section \ref{test}, does rely on calculating  NLS estimates of the parameters in $\boldsymbol{\theta}_0$ for $q=1$.  And, Theorem \ref{computation_thm} provides a formula for computing the NLS estimate of $\boldsymbol{\theta}_0$.  In particular, for $q=1$, Eq. (\ref{FWL_theta}) simplifies to
\begin{equation*} 
	  \widehat{\boldsymbol{\theta}} =  \left( (\dot{\boldsymbol{z}} \otimes \widehat{\boldsymbol{\Lambda}})'\boldsymbol{M}(\dot{\boldsymbol{z}} \otimes \widehat{\boldsymbol{\Lambda}})\right)^{-1}(\dot{\boldsymbol{z}} \otimes \widehat{\boldsymbol{\Lambda}})'\boldsymbol{M}\dot{\boldsymbol{y}}. 
\end{equation*}
with $\widehat{\boldsymbol{\Lambda}}' = (1, \widehat{\boldsymbol{\lambda}}') = (1, \widehat{\lambda}_2,\ldots,\widehat{\lambda}_T)$.  

\section{Monte Carlo experiments\label{monte}}
In this section, we evaluate the performance of the small-$T$ PIE estimator of $\boldsymbol{\beta}_0$ through Monte Carlo simulations of two models. In both studies we compare the estimator to the TWFE estimator and the large-$T$ IE estimator. For the large-$T$ estimator we include a bias correction term to account for serial correlation and conditional heteroskedasticity across $t$.  It is constructed as suggested in Remark 6 in \cite{Bai2009}.  We also apply the test statistic described in Section \ref{test}.

\subsection{Model 1}
We first considered an IE model with a scalar individual effect and a moderately persistent idiosyncratic error term:
\begin{align*}
y_{it}&=\beta_1x_{it1}+\beta_2x_{it2}+\delta_t+2\phi_t \widetilde{\eta}_i+1.4\epsilon_{it}\\
\epsilon_{it}&=0.8\epsilon_{it-1}+0.5\varepsilon_{it}
\end{align*}
where $\widetilde{\eta}_i,\epsilon_{i1},\varepsilon_{i1},\ldots, \varepsilon_{iT}\sim_{i.i.d.} N(0,1)$, $\phi_t=1-(t-1)/T$, and $\beta_1=-1$ and $\beta_2=1$. We generated the regressors according to:
\begin{align*}
x_{it1}&=\widetilde{\eta}_i+u_{it}^{x_1}\\
x_{it2}&=\phi_t \widetilde{\eta}_i+u_{it}^{x_2}
\end{align*}
where $u_{i1}^{x_1},u_{i1}^{x_2},\ldots, u_{iT}^{x_1},u_{iT}^{x_2}\sim_{i.i.d.} N(0,1)$, which were generated independently of $\widetilde{\eta}_i,\epsilon_{i1}$ and $\varepsilon_{i1},\ldots, \varepsilon_{iT}$. The first regressor, $x_{it1}$, does not have a time-varying covariance with $\widetilde{\eta}_i$ but $x_{it2}$ does. As a result, the TWFE estimator is consistent for $\beta_1$ but not for $\beta_2$.

\begin{figure}[h]
\begin{center}
\caption{Table 1: Monte Carlo results for Model 1}
\includegraphics[width=.85\textwidth]{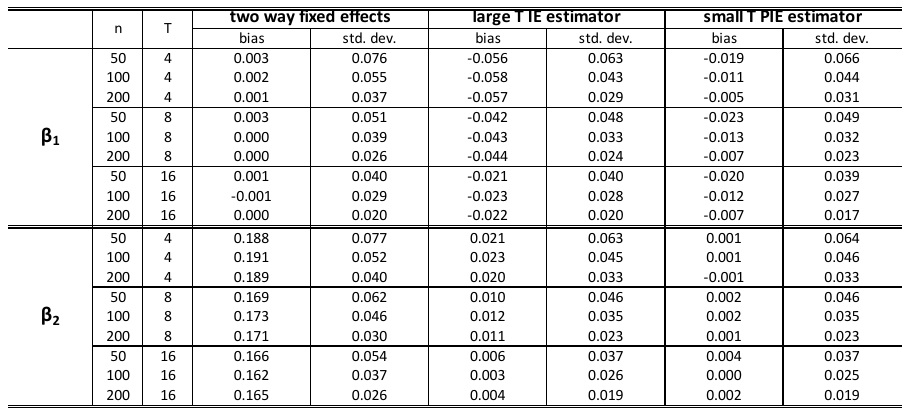}
\end{center}

\vspace{-20pt}
\hspace{30pt}{\scriptsize Notes: For each $n$ and $T$, we simulated the bias and standard deviation through a Monte Carlo exercise with}

\vspace{-5pt}
\hspace{30pt}{\scriptsize 1,000 iterations. }
\end{figure}

In Table 1 we report results for various values of $n$ and $T$. The results bear out what the asymptotic theory suggests. For  small $T$, the large-$T$ IE estimator exhibits bias that does not diminish as $n$ increases. In general, this bias is not in the same direction as the bias of the fixed effects estimator. However, as $T$ increases the large-$T$ IE estimator's bias becomes smaller. The small-$T$ PIE estimator, on the other hand, can have a non-negligible bias when $n$ is small, regardless of the size of $T$. In our simulations this bias is small in comparison to the bias of the fixed effects estimator and the large-$T$ IE estimator, though there is no theoretical guarantee that this will always be the case. Moreover, the PIE estimator's bias diminishes quickly as $n$ increases. The simulations also demonstrate the advantages of the small-$T$ PIE estimator do not necessarily come at the expense of higher variance. The PIE estimator's variance is generally similar in magnitude to the variance of the TWFE estimator.\footnote{In our unreported simulations, we also compared our small-$T$ estimator to the two-step GMM estimator studied in \cite{Ahn2001}. Neither of the two estimators dominates the other in terms of variance or small $n$ bias.}

We also examined the finite sample performance of the test statistic proposed in Section~\ref{test}.\footnote{As noted in Section \ref{test}, when constructing residuals for this test statistic, an estimator $\boldsymbol{b}$ of $\boldsymbol{\beta}_0$ is required, which can be either the PIE of TWFE estimator.  We used the TWFE estimator.} We modified the above model only by changing how $x_{it2}$ is generated. Specifically, for $s\in[0,1]$, we set $x_{it2}^{(s)}=(s\phi_t+(1-s))\widetilde{\eta}_i+u_{it}^{x_2}$. If $s=1$ then this produces the model used to generate the data in Table 1. On the other hand, setting $s=0$  produces a model that can be consistently estimated by the TWFE estimator because neither regressor has a time-varying covariance with $\widetilde{\eta}_i$. We simulated this model for several values of $s\in[0,1]$. The rejection probabilities are plotted in Figure 1 for simulations with $T$ set to four.

\begin{figure}[h]
\begin{center}
\caption{Figure 1: Rejection probability of Specification Test in Model 1 for $T=4$}
\includegraphics[width=.85\textwidth]{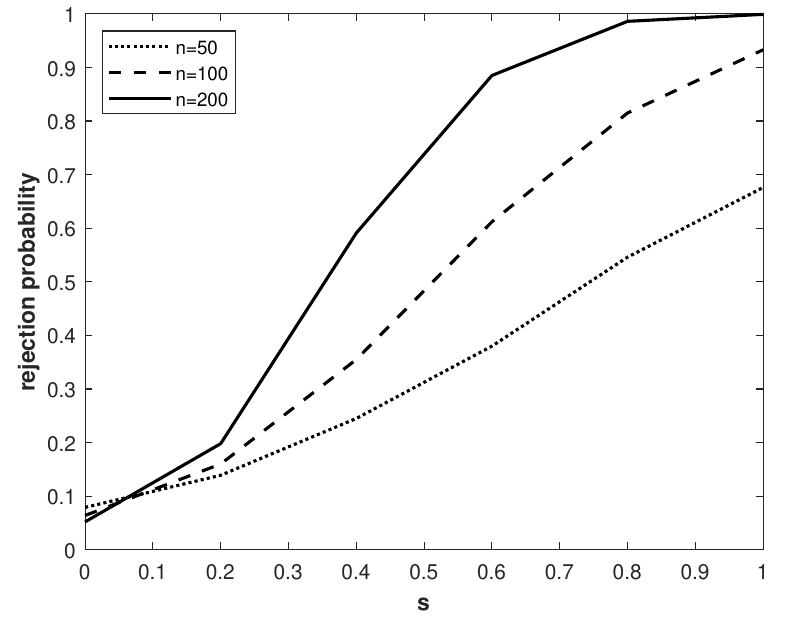}
\end{center}

\vspace{-30pt}
\hspace{35pt}{\scriptsize Notes: For each value of $s$ on a grid between $0$ and $1$, we simulated the power of the proposed test through}

\vspace{-5pt}
\hspace{35pt}{\scriptsize a Monte Carlo exercise with 1,000 iterations. }
\end{figure}

\subsection{Model 2}

Next, we examined a ``treatment effect'' model with staggered adoption.  For these experiments, we assumed a total of at most 16 observed periods.   For all $i$, we set the treatment variable $x_{it}=0$ for $t=1,\ldots,7$ and $x_{it}=1$ for $t=10,\ldots, 16$.  However, adoption of the policy (i.e., $x_{it} = 1$)---during either the $8$th, $9$th, or $10$th period---was randomly determined and correlated with the individual's individual-specific effect $\widetilde{\eta}_i$. Specifically, $\Pr(x_{i8}=1\mid \widetilde{\eta}_i)=0.93\widetilde{\eta}_i$, where $\widetilde{\eta}_i$ was either 0 or 1 with $\Pr(\widetilde{\eta}_i = 1) = 0.5$.  If $x_{i8}=1$, then we set $x_{i9}=x_{i10}=1$ as well. Moreover, $\Pr(x_{i9}=1\mid x_{i8}=0,\widetilde{\eta}_i)=0.69$, and if $x_{i9}=1$, then $x_{i10}=1$ as well.  

All 16 periods were not assumed to be observed in all of the experiments. In one set of experiments, we assumed only periods $7$ through $10$ were observed.  Therefore, for these experiments, a total of only $T=4$ periods were observed.  We also conducted experiments for which $T=8$ periods were observed.  For these experiments, periods  $5$ through $12$ were observed.  Finally, for the last set of experiments, estimation was based on data for all $T=16$ periods.

The outcome variable, $y_{it}$, was generated according to
\begin{equation*}
y_{it}=\delta_t+\beta_1x_{it}+\phi_t\widetilde{\eta}_i+e_{it} \qquad (t=t_1,\ldots, t_1+T-1, \; i= 1,\ldots, n),
\end{equation*}
where $T\in\{4,8,16\}$ and $t_1:=(16-T)/2+1$ is the first period observed, and $\beta_1=1$. The values for $\delta_t$ and $\phi_t$  are plotted in Figure 2.  As already noted, we set $\widetilde{\eta}_i = 0$ or 1 with $\Pr(\widetilde{\eta}_i = 1) = 0.5$.  Moreover, for each $i$, the errors were generated according to
\begin{align*}
e_{it_1}&=\varepsilon_{it_1}\\
e_{it}&=0.9e_{it-1}+\sqrt{1-0.9^2}\varepsilon_{it} \qquad (t=t_1+1,\ldots,t_1+T-1),
\end{align*}
where $\varepsilon_{i1},\ldots,\varepsilon_{i16}\sim_{i.i.d.}N(0,1)$ were generated independently from $\widetilde{\eta}_i,\{x_{it}\}_{t=1}^{16}$.  Finally, the values for $n$ that we considered were 50, 100, and 200.

\begin{figure}[h]
\begin{center}
\caption{Figure 2: Non-parallel trends}
\includegraphics[width=.6\textwidth]{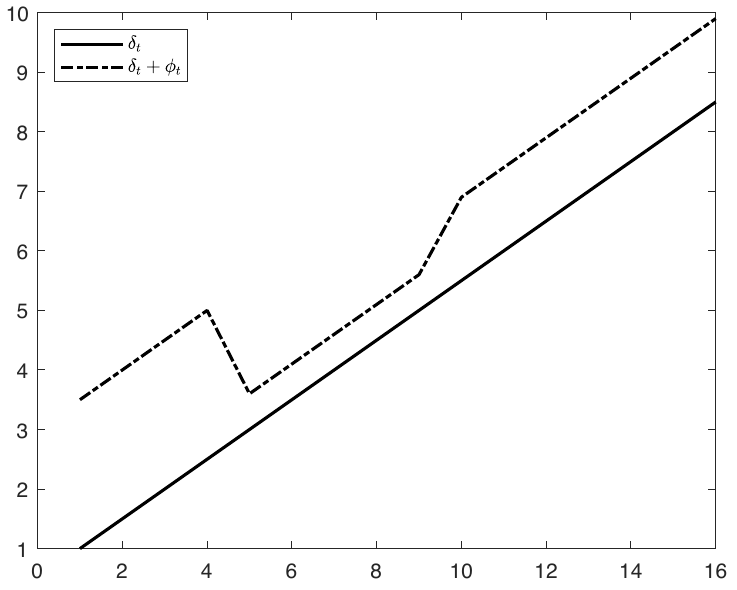}
\end{center}
\end{figure}

Table 2 summarizes the results from the Monte Carlo experiments. It is clear from the data in Table 2 that although the PIE estimator's standard deviation always exceeds that of both the TWFE and large-T IE estimators, the PIE estimator exhibits much less bias.  Consequently, it always improves on the TWFE and large-T IE estimators in terms of root mean squared error.  The median reduction in root mean squared error of the PIE estimator relative the TWFE estimator is about 56 percent, while the median reduction in its root mean squared error compared to the large-T IE estimator is about 59 percent.

\begin{figure}[h]
\begin{center}
\caption{Table 2: Monte Carlo results for Model 2}
\includegraphics[width=.85\textwidth]{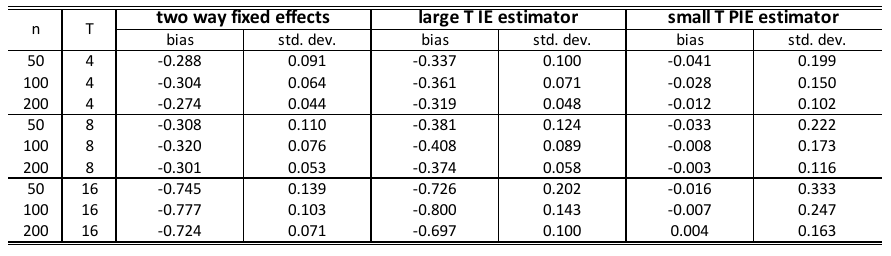}
\end{center}

\vspace{-20pt}
\hspace{30pt}{\scriptsize Notes: For each $n$ and $T$, we simulated the bias and standard deviation through a Monte Carlo exercise with}

\vspace{-5pt}
\hspace{30pt}{\scriptsize 1,000 iterations. }
\end{figure}

In order to investigate the size and power of the test described in Section \ref{test}, we also simulated the model while setting $\phi_t=s\phi_t^*+(1-s)\phi_{t_1}^*$ for $s\in[0,1]$; $\phi_t^*$ is as plotted in Figure 2. If $s=1$, we get the model used to generate the data in Table 2. If $s=0$, we get constant factor loadings, implying a TWFE model. Figure 3 shows the rejection probabilities for simulations where we assumed only periods 7 through 10 (i.e., $T=4$) are observed.

\begin{figure}[h]
\begin{center}
\caption{Figure 3: Rejection probability of Specification Test in Model 2 for $T=4$}
\includegraphics[width=.85\textwidth]{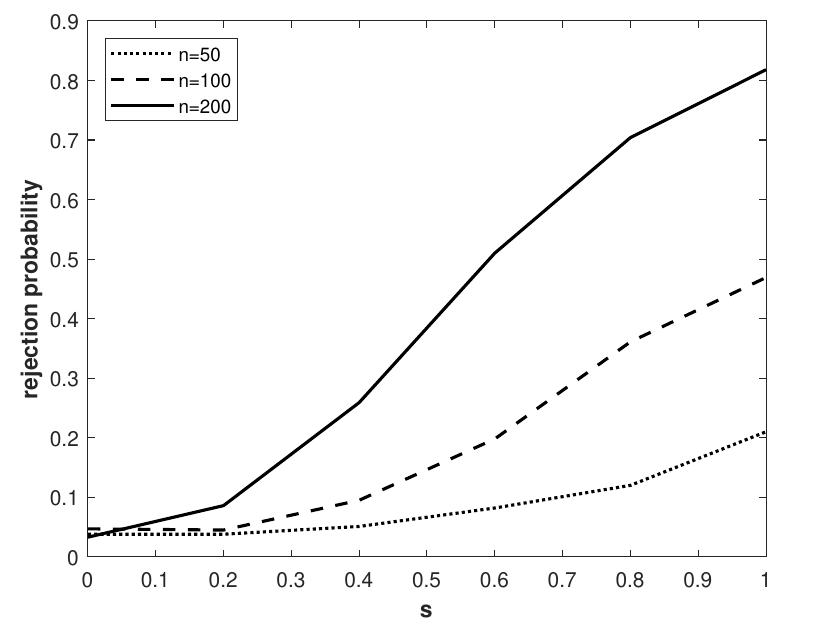}
\end{center}

\vspace{-30pt}
\hspace{35pt}{\scriptsize Notes: For each value of $s$ on a grid between $0$ and $1$, we simulated the power of the proposed test through}

\vspace{-5pt}
\hspace{35pt}{\scriptsize a Monte Carlo exercise with 1,000 iterations. }
\end{figure}

\section{Conclusion}
Although it is well-known that application of the within transformation and using a linear projection produce the same TWFE estimator in the conventional AE model, a similar result does not hold for the IE model. We apply the projection approach to the IE model to produce an estimator that is consistent for fixed $T$ in the presence of serially correlated and conditionally heteroskedastic errors. We show that this small-$T$ PIE estimator takes the form of a generalized within estimator, and we provide a simple iterative scheme for computing  estimates. We also derive a test of the null hypothesis that the AE model is sufficient. We show through Monte Carlo simulations that the estimator and test perform well under a variety of circumstances where the TWFE estimator does not, and that it outperforms the large-$T$ IE estimator of \cite{Bai2009}. 

\newpage
\section*{Appendix A: Proofs \label{proofs}}

All limits are taken with $ n \rightarrow \infty $ and $ T $ fixed.

\subsection*{A.1 $\;$ Theorem \ref{consistency_thm} proof}

 Recall that $ \boldsymbol{y}_{i}^{\ast} := \boldsymbol{y}_i - E\left( \boldsymbol{y}_1\right)  $ and  $ \boldsymbol{X}_i^{\ast} $, $ \boldsymbol{Z}_i^{\ast} $ are defined similarly. Moreover, let $ \boldsymbol{v}_i^{\ast} := \boldsymbol{v}_i - E\left( \boldsymbol{v}_1\right) = (\boldsymbol{z}_i^{
 \ast \prime}, \boldsymbol{a}_i', \boldsymbol{\epsilon}_i')'  $. Set 
 \begin{equation} \label{s_star}
 	\begin{split}
 s(\boldsymbol{v}_i^{\ast}, \boldsymbol{\psi}) &:= \frac{1}{2}\left( \boldsymbol{y}_i^{\ast} - \boldsymbol{X}_i^{\ast}\boldsymbol{\beta} - \boldsymbol{\Lambda}\boldsymbol{Z}_i^{\ast}\boldsymbol{\theta}\right) '\left( \boldsymbol{y}_i^{\ast} - \boldsymbol{X}_i^{\ast}\boldsymbol{\beta} - \boldsymbol{\Lambda}\boldsymbol{Z}_i^{\ast}\boldsymbol{\theta}\right) \\
 & =\frac{1}{2}\sum_t(y_{it}^{\ast} - \boldsymbol{z}_i^{\ast \prime}\boldsymbol{\pi}_t(\boldsymbol{\psi}))^2. \\
 \end{split}
\end{equation}  
Also, let $ \overline{S}(\boldsymbol{\psi}) := E\left[ s(\boldsymbol{v}_1^{\ast},\boldsymbol{\psi})\right] $.  The theorem is proven upon showing that $ S_{n}( \cdot) \rightarrow_p \overline{S}(\cdot) $  uniformly on $ \overline{\boldsymbol{\Psi}} $, and $ \overline{S}(\boldsymbol{\psi}_0) < \overline{S}(\boldsymbol{\psi})$ for $ \boldsymbol{\psi} \neq \boldsymbol{\psi}_0$ and $ \boldsymbol{\psi} \in \overline{\boldsymbol{\Psi}} $, where  $ \overline{\boldsymbol{\Psi}} $ is  compact  with $ \boldsymbol{\psi}_0 \in \overline{\boldsymbol{\Psi}} $.  Given these conditions, it follows that $ \widehat{\boldsymbol{\psi}} \rightarrow_p \boldsymbol{\psi}_0 $ (see, e.g., Newey and McFadden, 1994, p. 2121, Theorem 2.1).

 We first show that $\boldsymbol{\psi}_0$ is identified in the sense that $ \overline{S}(\boldsymbol{\psi}_0) < \overline{S}(\boldsymbol{\psi})$ for $ \boldsymbol{\psi} \neq \boldsymbol{\psi}_0$ and $ \boldsymbol{\psi} \in \boldsymbol{\Psi} $. If A1 through A4 are satisfied, then A5 is  necessary and sufficient for identification of $\boldsymbol{\psi}_0$.
 
 {\em Sufficiency}: Note that
 \begin{equation*}
 \begin{split}
2s(\boldsymbol{v}_1^{\ast},\boldsymbol{\psi}) 
& = \sum_t\left(y_{1t}^{\ast} - \boldsymbol{z}_1^{\ast \prime}\boldsymbol{\pi}_t(\boldsymbol{\psi})\right)^2  \\
& = \sum_t\left[u_{1,t} + \boldsymbol{z}_1^{\ast \prime}\left(\boldsymbol{\pi}_t(\boldsymbol{\psi}_0)- \boldsymbol{\pi}_t(\boldsymbol{\psi})\right)\right]^2 \\
& = \sum_t u_{1,t}^2 + 2\sum_tu_{1,t}\boldsymbol{z}_1^{\ast \prime}\left(\boldsymbol{\pi}_t(\boldsymbol{\psi}_0)- \boldsymbol{\pi}_t(\boldsymbol{\psi})\right) \\
& + \sum_t\left(\boldsymbol{\pi}_t(\boldsymbol{\psi}_0)- \boldsymbol{\pi}_t(\boldsymbol{\psi})\right)'\boldsymbol{z}_1^{\ast}\boldsymbol{z}_1^{\ast \prime}\left(\boldsymbol{\pi}_t(\boldsymbol{\psi}_0)- \boldsymbol{\pi}_t(\boldsymbol{\psi})\right).
\end{split}
\end{equation*}
It follows from $ E(\boldsymbol{z}_1 \boldsymbol{a}_1^{\prime}) = \boldsymbol{0} $ and Assumption A3 that $ E(u_{1t}\boldsymbol{z}_1^{\ast} ) = \boldsymbol{0} $ for all $t$.   Hence,  
\begin{equation} \label{identification_result}
2\overline{S}(\boldsymbol{\psi})  =  \sum_tE\left(u_{1,t}^2\right) 
 + \sum_t\left(\boldsymbol{\pi}_t(\boldsymbol{\psi}_0)- \boldsymbol{\pi}_t(\boldsymbol{\psi})\right)'\text{Var}(\boldsymbol{z}_1)\left(\boldsymbol{\pi}_t(\boldsymbol{\psi}_0)- \boldsymbol{\pi}_t(\boldsymbol{\psi})\right).
\end{equation}
From this last expression we see that Assumptions A1 through A4 are satisfied, then A5 implies $ \overline{S}(\boldsymbol{\psi}_0) < \overline{S}(\boldsymbol{\psi})$ for $ \boldsymbol{\psi} \neq \boldsymbol{\psi}_0$ and $ \boldsymbol{\psi} \in \boldsymbol{\Psi} $.

{\em Necessity}: From (\ref{identification_result}) we see that $ \overline{S}(\boldsymbol{\psi}_0) < \overline{S}(\boldsymbol{\psi})$ for $ \boldsymbol{\psi} \neq \boldsymbol{\psi}_0$ and $ \boldsymbol{\psi} \in \boldsymbol{\Psi} $ implies that  for each $\boldsymbol{\psi} \in \boldsymbol{\Psi}$ such that $\boldsymbol{\psi} \ne \boldsymbol{\psi}_0$, there is at least one $t \in \{1,\ldots,T\}$ such that $\boldsymbol{\pi}_t(\boldsymbol{\psi}) \ne \boldsymbol{\pi}_t(\boldsymbol{\psi}_0)$. 
 
 Given the function $\overline{S}(\cdot)$ is uniquely minimized at $\boldsymbol{\psi}_0$, there is a measurable minimizer, say $ \widehat{\boldsymbol{\psi}} $, of $ S_n(\cdot) $ in $ \overline{\boldsymbol{\Psi}} $ such that $ \widehat{\boldsymbol{\psi}} \rightarrow_p \boldsymbol{\psi}_0 $ if $ S_{n}(\cdot) \rightarrow_p \overline{S}(\cdot) $ uniformly on $ \overline{\boldsymbol{\Psi}} $ (see, e.g., Newey and McFadden, 1994, p. 2121, Theorem 2.1).  It remains to show that $ S_{n}(\cdot) \rightarrow_p \overline{S}(\cdot) $ uniformly on $ \overline{\boldsymbol{\Psi}} $.

 To establish this conclusion, define $ S_{n}^{\ast}(\boldsymbol{\psi}) := (1/n)\sum_i s(\boldsymbol{v}_i^{\ast}, \boldsymbol{\psi}) $.  Then $ S_{n}( \boldsymbol{\psi}) - S_{n}^{\ast}( \boldsymbol{\psi}) = - s(\overline{\boldsymbol{v}}^{\ast},\boldsymbol{\psi}) $, where $s(\overline{\boldsymbol{v}}^{\ast},\boldsymbol{\psi}) := \sum_t [\overline{y}_t - E(y_{1t}) - (\overline{\boldsymbol{z}} - E(\boldsymbol{z}_1) )'\boldsymbol{\pi}(\boldsymbol{\psi}) ]^2/2$. By the c$_{r}$-inequality \citep[see, e.g.,][p. 21]{Sen1993}, we have 
 \begin{equation*}
 s(\overline{\boldsymbol{v}}^{\ast},\boldsymbol{\psi})  \le \sum_t \{[\overline{y}_t - E(y_{1t})]^2 + (\overline{\boldsymbol{z}} - E(\boldsymbol{z}_1) )'\boldsymbol{\pi}(\boldsymbol{\psi})\boldsymbol{\pi}(\boldsymbol{\psi})' (\overline{\boldsymbol{z}} - E(\boldsymbol{z}_1) )\}
\end{equation*}
Under our assumptions,  $\sum_t [\overline{y}_t - E(y_{1t})]^2 \rightarrow_p 0$ independently of $\boldsymbol{\psi} \in \boldsymbol{\Psi}$ and thus uniformly on $\boldsymbol{\Psi}$.  Let $\overline{z}_j$ and $E(z_{1j})$ denote the $j$th entries of $\overline{\boldsymbol{z}}$ and $E(\boldsymbol{z}_1)$, and let $\varphi_{jk}(\boldsymbol{\psi})$ denote the $(j,k)$th entry of $\boldsymbol{\pi}(\boldsymbol{\psi})\boldsymbol{\pi}(\boldsymbol{\psi})'$. Then
\begin{equation*}
\sum_t  \left(\overline{\boldsymbol{z}} - E(\boldsymbol{z}_1) \right)'\boldsymbol{\pi}(\boldsymbol{\psi})\boldsymbol{\pi}(\boldsymbol{\psi})' \left(\overline{\boldsymbol{z}} - E(\boldsymbol{z}_1) \right)  = \sum_t \sum_j \sum_k
 \varphi_{jk}(\boldsymbol{\psi})(\overline{z}_j - E(z_{1j}) )(\overline{z}_k - E(z_{1k}) ) 
\end{equation*}
Moreover, the right-hand side of the last expression is bounded above by
\begin{equation} \label{bound}
\begin{split}
 \sum_t \sum_j \sum_k & \left|
 \varphi_{jk}(\boldsymbol{\psi})\right| \left|(\overline{z}_j   - E(z_{1j}) )(\overline{z}_k - E(z_{1k}) )\right|  \\
 & \le M \sum_t \sum_j \sum_k  \left|(\overline{z}_j - E(z_{1j}) )(\overline{z}_k - E(z_{1k}) )\right| \quad \forall \; \boldsymbol{\psi} \in \overline{\boldsymbol{\Psi}},
 \end{split}
\end{equation}
for some finite positive number $M$. The inequality in (\ref{bound}) follows from the fact that, fall all $j$ and $k$, the function $\varphi_{jk}(\cdot)$ is continuous on the compact set $\overline{\boldsymbol{\Psi}}$ and is, therefore, bounded. Given $\overline{z}_j \rightarrow_p E(z_{1j})$ ($j = 1,\ldots,m$), the right-hand side of (\ref{bound}) converges in probability to zero independently of $\boldsymbol{\psi} \in \overline{\boldsymbol{\Psi}}$ and thus uniformly on $\overline{\boldsymbol{\Psi}}$.  The proceeding shows that $ S_{n}( \cdot) - S_{n}^{\ast}( \cdot) \rightarrow_p 0 $ uniformly on $\overline{\boldsymbol{\Psi}}$.

 The conclusion of the last paragraph implies $ S_{n}( \cdot)  \rightarrow_p  \overline{S}(\cdot) $ uniformly on $ \overline{\boldsymbol{\Psi}} $ if $ S_{n}^{\ast}( \cdot) \rightarrow_p  \overline{S}(\cdot) $ uniformly on $ \overline{\boldsymbol{\Psi}} $.  
 In order to establish the uniform convergence of $S_{n}^{\ast}(\cdot) $ it suffices to find a function $ b(\boldsymbol{v}_i^{\ast}) $ such that $ s(\boldsymbol{v}_i^{\ast},\boldsymbol{\psi}) \leq b(\boldsymbol{v}_i^{\ast}) $, for all $ \boldsymbol{\psi} \in \overline{\boldsymbol{\Psi}} $, and $ E\left[ b(\boldsymbol{v}_1^{\ast}) \right] < \infty $ (see Newey and McFadden, 1994, p. 2129, Lemma 2.4). 
 
 To that end, let $z_{ij}^{\ast}$ denote the $j$th entry in $\boldsymbol{z}_i^{\ast}$. Then
 \begin{equation} \label{bound2}
 \begin{split}
s(\boldsymbol{v}_i^{\ast},\boldsymbol{\psi}) 
& = \frac{1}{2}\sum_t\left(y_{it}^{\ast} - \boldsymbol{z}_i^{\ast \prime}\boldsymbol{\pi}_t(\boldsymbol{\psi})\right)^2  \\
& \le  \sum_t\left(y_{it}^{\ast 2} + \boldsymbol{z}_i^{\ast \prime}\boldsymbol{\pi}_t(\boldsymbol{\psi})\boldsymbol{\pi}_t(\boldsymbol{\psi})'\boldsymbol{z}_i^{\ast \prime} \right) \\
& \le M \sum_t\left(y_{it}^{\ast 2} + \sum_j \sum_k z_{ij}^{\ast }z_{ik}^{\ast} \right) \qquad \forall \; \boldsymbol{\psi} \in \overline{\boldsymbol{\Psi}},
\end{split}
\end{equation}
 where the first inequality follows from the c$_r$-inequality and the second inequality follows from an argument similar to that used to establish (\ref{bound}).  Let $ b(\boldsymbol{v}_i^{\ast})$ denote the right-hand side of (\ref{bound2}).  Our assumptions ensure $ E\left[ b(\boldsymbol{v}_1^{\ast}) \right] < \infty $, which completes the proof.

\subsection*{A.2 $\;$ Theorem \ref{normality_thm} proof}

\sloppy
Let $\boldsymbol{g}_n(\boldsymbol{\psi}) := \partial S_n(\boldsymbol{\psi})/\partial \boldsymbol{\psi}$ and $ \boldsymbol{H}_n(\boldsymbol{\psi}) := \partial^2 S_n(\boldsymbol{\psi})/\partial \boldsymbol{\psi} \partial \boldsymbol{\psi}^{\prime} $. Given $ \widehat{\boldsymbol{\psi}} \rightarrow_p \boldsymbol{\psi}_0 $ and $ \boldsymbol{H}(\boldsymbol{\psi}) :=  \partial^2 E\left[ s(\boldsymbol{v}_1^{\ast},\boldsymbol{\psi}) \right]/\partial \boldsymbol{\psi} \partial \boldsymbol{\psi}^{\prime} $ is continuous in $ \boldsymbol{\psi} $, it suffices to show that $ \sqrt{n} \, \boldsymbol{g}_n(\boldsymbol{\psi}_0) \rightarrow_d N(0,\boldsymbol{A}_0)$; $ \boldsymbol{H}_n(\cdot) \rightarrow_p \boldsymbol{H}(\cdot)$ uniformly on $ \overline{\boldsymbol{\Psi}} $; and $ \boldsymbol{H}_0 = \boldsymbol{H}(\boldsymbol{\psi}_0) $ is nonsingular (Newey and McFadden, 1994, p. 2143, Theorem 3.1). 

Let $\boldsymbol{g}_n^{\ast}(\boldsymbol{\psi}) := \partial S_n^{\ast}(\boldsymbol{\psi})/\partial \boldsymbol{\psi}$, and note that $ \sqrt{n} \, \boldsymbol{g}_n(\boldsymbol{\psi}_0) = \sqrt{n} \, \boldsymbol{g}_n^{\ast}(\boldsymbol{\psi}_0)  -  \sqrt{n} \, \partial s(\overline{\boldsymbol{v}}^{\ast},\boldsymbol{\psi}_0)/\partial \boldsymbol{\psi} $.  Calculations show that $ \partial s(\boldsymbol{v}_i^{\ast}, \boldsymbol{\psi}_0)/\partial \boldsymbol{\beta}  =  -  \boldsymbol{X}_{i}^{\ast \prime}\boldsymbol{u}_i$, $ \partial s(\boldsymbol{v}_i^{\ast}, \boldsymbol{\psi}_0)/\partial \boldsymbol{\theta}  =  -  \boldsymbol{Z}_{i}^{\ast \prime}\boldsymbol{\Lambda}_0'\boldsymbol{u}_i$, and $ \partial s(\boldsymbol{v}_i^{\ast}, \boldsymbol{\psi}_0)/\partial \boldsymbol{\lambda}_j  = -  \boldsymbol{z}_i^{\ast \prime}\boldsymbol{\theta}_{0j} \boldsymbol{u}_{2i}  $ ($j=1,\ldots, q$), where $\boldsymbol{\lambda}_j = (\lambda_{q+1,j},\ldots,\lambda_{Tj})'$ is the $j$th column of $\boldsymbol{\Lambda}_2$, and $ \boldsymbol{u}_{2i} $ is the last $ T-q $ rows of $ \boldsymbol{u}_{i} $.  Hence, $ \sqrt{n} \, \boldsymbol{g}_n^{\ast}(\boldsymbol{\psi}_0)  = - (1/\sqrt{n})\sum_{i} \boldsymbol{R}_i^{\ast \prime}\boldsymbol{u}_i  $.  Moreover, $ \sqrt{n} \partial s(\overline{\boldsymbol{v}}^{\ast},\boldsymbol{\psi}_0)/\partial \boldsymbol{\psi} =  - \overline{\boldsymbol{R}}^{\ast \prime}\sqrt{n}\overline{\boldsymbol{u}} $, where $ \overline{\boldsymbol{R}}^{\ast} := (1/n)\sum_{i} \boldsymbol{R}_i^{\ast}$. By the law of large numbers,  $ \overline{\boldsymbol{R}}^{\ast} \rightarrow_p \boldsymbol{0} $. And, Assumptions A1 and A2 imply $ \text{Var}(\sqrt{n}\overline{\boldsymbol{u}}) = \text{Var}(\boldsymbol{u}_1) $, which implies $ \sqrt{n}\overline{\boldsymbol{u}} = O_{p}(1) $.  Hence, $ \sqrt{n} \, \partial s(\overline{\boldsymbol{v}}^{\ast},\boldsymbol{\psi}_0)/\partial \boldsymbol{\psi} =  o_{p}(1) $.  Moreover, by the central limit theorem for independently and identically distributed random vectors (see, e.g., Greenberg and Webster, 1983, p. 21, Theorem 1.2.4), it follows that $   - (1/\sqrt{n})\sum_{i} \boldsymbol{R}_i^{\ast \prime}\boldsymbol{u}_i  \rightarrow_d N(\boldsymbol{0},\boldsymbol{A}_0)$, where $ \boldsymbol{A}_0 = E\left( \boldsymbol{R}_i^{\ast \prime}\boldsymbol{u}_i \boldsymbol{u}_i^{\prime} \boldsymbol{R}_i^{\ast} \right)  $ has finite entries given Assumption A2$^{\ast}  $.  The preceding verifies $  \sqrt{n} \, \boldsymbol{g}_n(\boldsymbol{\psi}_0)  \rightarrow_d N(\boldsymbol{0},\boldsymbol{A}_0)$.

\fussy
Let $ \boldsymbol{H}_n^{\ast}(\boldsymbol{\psi}) := \partial^{2}S_{n}^{\ast}(\boldsymbol{\psi})/\partial\boldsymbol{\psi}\partial\boldsymbol{\psi}^{\prime}$. Note that $ \boldsymbol{H}_n(\boldsymbol{\psi})  =  \boldsymbol{H}_n^{\ast}(\boldsymbol{\psi}) - \partial^2 s(\overline{\boldsymbol{v}}^{\ast},\boldsymbol{\psi})/\partial \boldsymbol{\psi} \partial \boldsymbol{\psi}' $.  The entries in  $ \partial^2 s(\overline{\boldsymbol{v}}^{\ast},\boldsymbol{\psi})/\partial \boldsymbol{\psi} \partial \boldsymbol{\psi}' $ are given by
\begin{align*}
	\frac{\partial^2 s(\overline{\boldsymbol{v}}^{\ast},\boldsymbol{\psi})}{\partial \boldsymbol{\gamma} \partial \boldsymbol{\gamma}^{\prime}} & = 
\left(
\begin{array}{cc}
  \overline{\boldsymbol{X}}^{\ast \prime}\overline{\boldsymbol{X}}^{\ast} & \overline{\boldsymbol{X}}^{\ast \prime} \boldsymbol{\Lambda} \overline{\boldsymbol{Z}}^{\ast}    \\
\overline{\boldsymbol{Z}}^{\ast \prime} \boldsymbol{\Lambda}' \overline{\boldsymbol{X}}^{\ast}   &     \overline{\boldsymbol{Z}}^{\ast \prime} \boldsymbol{\Lambda}'\boldsymbol{\Lambda} \overline{\boldsymbol{Z}}^{\ast}
\end{array}
\right),
 \\
	\frac{\partial^2 s(\overline{\boldsymbol{v}}^{\ast},\boldsymbol{\psi})}{\partial \boldsymbol{\lambda} \partial \boldsymbol{\beta}^{\prime}} & =  \left( \overline{\boldsymbol{Z}}^{\ast} \boldsymbol{\theta} \otimes \boldsymbol{I}_{T-q} \right)  \overline{\boldsymbol{X}}_{2}^{\ast}, \\
	\frac{\partial^2 s(\overline{\boldsymbol{v}}^{\ast},\boldsymbol{\psi})}{\partial \boldsymbol{\lambda}_k \partial \boldsymbol{\theta}_k^{\prime}} & = - \left( \overline{\boldsymbol{y}}_{2}^{\ast} - \overline{\boldsymbol{X}}_{2}^{\ast}\boldsymbol{\beta} - \boldsymbol{\Lambda}_2\overline{\boldsymbol{Z}}^{\ast}\boldsymbol{\theta} \right) \overline{\boldsymbol{z}}^{\ast \prime} + \overline{\boldsymbol{z}}^{\ast \prime} \boldsymbol{\theta}_k \boldsymbol{\lambda}_k\overline{\boldsymbol{z}}^{\ast \prime},
\end{align*}
where $ \overline{\boldsymbol{y}}_{2}^{\ast} $, and $ \overline{\boldsymbol{X}}_{2}^{\ast} $ are the last $ T-q $ rows of $ \overline{\boldsymbol{y}}^{\ast} $ and $ \overline{\boldsymbol{X}}^{\ast} $, and, finally, 
\begin{align*}
	\frac{\partial^2 s( \overline{\boldsymbol{v}}^{\ast},\boldsymbol{\psi})}{\partial \boldsymbol{\lambda}_j \partial \boldsymbol{\theta}_k^{\prime}} & = \overline{\boldsymbol{z}}^{\ast \prime} \boldsymbol{\theta}_j \boldsymbol{\lambda}_k\overline{\boldsymbol{z}}^{\ast \prime} \qquad (j \neq k), \\
	\frac{\partial^2 s(\overline{\boldsymbol{v}}^{\ast},\boldsymbol{\psi})}{\partial \boldsymbol{\lambda}_j \partial \boldsymbol{\lambda}_k^{\prime}} & = \overline{\boldsymbol{z}}^{\ast \prime} \boldsymbol{\theta}_j \overline{\boldsymbol{z}}^{\ast \prime} \boldsymbol{\theta}_k \boldsymbol{I}_{T-q} .
\end{align*}
These partial derivatives reveal that each entry in $ \partial^2 s(\overline{\boldsymbol{v}}^{\ast},\boldsymbol{\psi})/\partial \boldsymbol{\psi} \partial \boldsymbol{\psi}' $ is a sum of products of two entries in $ \overline{\boldsymbol{v}}^{\ast} $ and entries in $ \boldsymbol{\psi} $. Hence, for $ \boldsymbol{\psi} \in \overline{\boldsymbol{\Psi}} $, there is a $ M < \infty $ such that $ \sup_{\boldsymbol{\psi} \in \overline{\boldsymbol{\Psi}}} |\partial^2 s(\overline{\boldsymbol{v}}^{\ast},\boldsymbol{\psi})/\partial \psi_l \partial \psi_m| \le M \sum_{j,k} |\overline{v}_j^{\ast}\overline{v}_k^{\ast}| $, where $\overline{v}_{j}^{\ast}$ and $\overline{v}_{k}^{\ast}$ denote the $j$th and $k$th entries in $\overline{\boldsymbol{v}}^{\ast}$. Moreover, $ \sum_{j,k} |\overline{v}_j^{\ast}\overline{v}_k^{\ast}| \rightarrow_p 0 $ because $ \overline{\boldsymbol{v}}^{\ast} \rightarrow_p \boldsymbol{0} $ by the law of large numbers. Hence, $ \boldsymbol{H}_n(\cdot) -  \boldsymbol{H}_n^{\ast}(\cdot) \rightarrow_p \boldsymbol{0} $ uniformly on  $ \overline{\boldsymbol{\Psi}} $.

\sloppy

Note that $ \boldsymbol{H}_n^{\ast}(\boldsymbol{\psi}) = (1/n)\sum_i \partial^2 s(\boldsymbol{v}_i^{\ast},\boldsymbol{\psi})/\partial\boldsymbol{\psi}\partial\boldsymbol{\psi}^{\prime} $.  Moreover, 
\begin{align*}
	 \frac{\partial^2 s(\boldsymbol{v}_i^{\ast},\boldsymbol{\psi})}{\partial \boldsymbol{\gamma} \partial \boldsymbol{\gamma}^{\prime}} & =    \left(
\begin{array}{cc}
  \boldsymbol{X}_i^{\ast \prime}\boldsymbol{X}_i^{\ast} & \boldsymbol{X}_i^{\ast \prime} \boldsymbol{\Lambda} \boldsymbol{Z}_i^{\ast}    \\
\boldsymbol{Z}_i^{\ast \prime} \boldsymbol{\Lambda}' \boldsymbol{X}_i^{\ast}   &     \boldsymbol{Z}_i^{\ast \prime} \boldsymbol{\Lambda}'\boldsymbol{\Lambda} \boldsymbol{Z}_i^{\ast}
\end{array}
\right),\\
	 \frac{\partial^2 s(\boldsymbol{v}_i^{\ast},\boldsymbol{\psi})}{\partial \boldsymbol{\lambda} \partial \boldsymbol{\beta}^{\prime}} & =   \left( \boldsymbol{Z}_i^{\ast} \boldsymbol{\theta} \otimes \boldsymbol{I}_{T-q} \right)  \boldsymbol{X}_{2i}^{\ast} , \\
	 \frac{\partial^2 s(\boldsymbol{v}_i^{\ast},\boldsymbol{\psi})}{\partial \boldsymbol{\lambda}_k \partial \boldsymbol{\theta}_k^{\prime}}  &=  -  \left( \boldsymbol{y}_{2i}^{\ast} - \boldsymbol{X}_{2i}^{\ast}\boldsymbol{\beta} - \boldsymbol{\Lambda}_2\boldsymbol{Z}_i^{\ast}\boldsymbol{\theta}\right) \boldsymbol{z}_i^{\ast \prime} + \boldsymbol{z}_i^{\ast \prime} \boldsymbol{\theta}_k \boldsymbol{\lambda}_k\boldsymbol{z}_i^{\ast \prime} , 
\end{align*}
where $ \boldsymbol{y}_{2i}^{\ast} $ and $ \boldsymbol{X}_{2i}^{\ast} $ are the last $ T-q $ rows of $ \boldsymbol{y}_{i}^{\ast} $ and $ \boldsymbol{X}_{i}^{\ast} $, and 
\begin{align*}
	\frac{\partial^2 s(\boldsymbol{v}_i^{\ast},\boldsymbol{\psi})}{\partial \boldsymbol{\lambda}_j \partial \boldsymbol{\theta}_k^{\prime}} & =  \boldsymbol{z}_i^{\ast \prime} \boldsymbol{\theta}_j \boldsymbol{\lambda}_k\boldsymbol{z}_i^{\ast \prime} \qquad (j \neq k), \\
	\frac{\partial^2 s(\boldsymbol{v}_i^{\ast},\boldsymbol{\psi})}{\partial \boldsymbol{\lambda}_j \partial \boldsymbol{\lambda}_k^{\prime}} & = \boldsymbol{z}_i^{\ast \prime} \boldsymbol{\theta}_j \boldsymbol{z}_i^{\ast \prime} \boldsymbol{\theta}_k \boldsymbol{I}_{T-q}. 
\end{align*}
It is clear from these partial derivatives that each entry in $ \partial^2 s(\boldsymbol{v}_i^{\ast},\boldsymbol{\psi})/\partial \boldsymbol{\psi} \partial \boldsymbol{\psi}^{\prime}$ is a sum, and the terms in these sums involve products of two entries in $\boldsymbol{v}_i^{\ast}$.  Furthermore, for many of the of these terms, the products of entries in $\boldsymbol{v}_i^{\ast}$ are multiplied by an entry or product of entries in $\boldsymbol{\psi}$. Because the entries in $\boldsymbol{\psi}$ are bounded for $\boldsymbol{\psi} \in \overline{\boldsymbol{\Psi}}$, it follows that there is a finite $M$ such that the absolute value of every entry in $ \partial^2 s(\boldsymbol{v}_i^{\ast},\boldsymbol{\psi})/\partial \boldsymbol{\psi} \partial \boldsymbol{\psi}^{\prime}$ is bounded by a sum of the form $ M \sum_{j,k} |v_{ij}^{\ast}v_{ik}^{\ast}| $, for all $\boldsymbol{\psi} \in \overline{\boldsymbol{\Psi}}$, where $v_{ij}^{\ast}$ and $v_{ik}^{\ast}$ denote the $j$th and $k$th entries in $\boldsymbol{v}_i^{\ast}$. By the Cauchy-Schwarz inequality, we have $ E|v_{1j}^{\ast}v_{1k}^{\ast}| \leq \left[ E(v_{1j}^{\ast 2})E(v_{1k}^{\ast 2}) \right]^{1/2} $,  and our assumptions ensure the right-hand side of this inequality is finite. Hence, the absolute value of every entry in $ \partial^2 s(\boldsymbol{v}_i^{\ast},\boldsymbol{\psi})/\partial \boldsymbol{\psi} \partial \boldsymbol{\psi}^{\prime}$ is no greater than some $b(\boldsymbol{v}_i^{\ast})$, for all $\boldsymbol{\psi} \in \overline{\boldsymbol{\Psi}}$, and $ E\left[ b(\boldsymbol{v}_1^{\ast})\right] < \infty $. Hence, $ \boldsymbol{H}_n^{\ast}(\cdot)  \rightarrow_p E\left[ \partial^2 s(\boldsymbol{v}_1,\cdot)/\partial \boldsymbol{\psi} \partial \boldsymbol{\psi}^{\prime}\right] $ uniformly on $\overline{\boldsymbol{\Psi}}$ (Newey and McFadden, 1994, p. 2129, Lemma 2.4). Moreover, $ E\left[ \partial^2 s(\boldsymbol{v}_1^{\ast},\boldsymbol{\psi})/\partial \boldsymbol{\psi} \partial \boldsymbol{\psi}^{\prime}\right] =  \partial^2 E\left[s(\boldsymbol{v}_1^{\ast},\boldsymbol{\psi})\right]/\partial \boldsymbol{\psi} \partial \boldsymbol{\psi}^{\prime} =  \partial^2 \overline{S}(\boldsymbol{\psi})/\partial \boldsymbol{\psi} \partial \boldsymbol{\psi}^{\prime}$. And $ \partial^2 \overline{S}(\boldsymbol{\psi}_0)/\partial \boldsymbol{\psi} \partial \boldsymbol{\psi}^{\prime} = E(\boldsymbol{R}_i^{\ast \prime}\boldsymbol{R}_i^{\ast}) = \boldsymbol{H}_0$ is positive definite given $\overline{S}(\cdot)$ is uniquely minimized at $\boldsymbol{\psi}_0$. This completes the proof.

\subsection*{A.3 $\;$ Theorem \ref{dist_4_test_stat} proof}

By the mean value theorem, 
\begin{equation} \label{mean_value}
\sqrt{n}\,\boldsymbol{g}_n(\widehat{\boldsymbol{\psi}}) = \sqrt{n}\,\boldsymbol{g}_n(\boldsymbol{\psi}_0) + \overline{\boldsymbol{H}}_n \sqrt{n}\left(\widehat{\boldsymbol{\psi}} - \boldsymbol{\psi}_0\right).
\end{equation}
In (\ref{mean_value}), $\boldsymbol{g}_n(\boldsymbol{\psi}) := \partial S_n(\boldsymbol{\psi})/\partial \boldsymbol{\psi}$ and $\overline{\boldsymbol{H}}_n$ is $p\times p$ matrix whose $(j,k)$th entry is the second-order partial derivative $\partial^2 S_n(\overline{\boldsymbol{\psi}}_{(j)})/\partial \psi_j \partial \psi_k$, where $\psi_j$ denotes the $j$th entry in $\boldsymbol{\psi}$ and $\overline{\boldsymbol{\psi}}_{(j)}$ is a vector that satisfies $\parallel \overline{\boldsymbol{\psi}}_{(j)} - \boldsymbol{\psi}_0 \parallel \le \parallel \widehat{\boldsymbol{\psi}} - \boldsymbol{\psi}_0 \parallel$ ($j=1,\ldots,p$).
It follows from (\ref{mean_value}) and  $\boldsymbol{g}_n(\widehat{\boldsymbol{\psi}}) = \boldsymbol{0}$ that
\begin{equation} \label{stack1}
\sqrt{n}\left(\widehat{\boldsymbol{\psi}} - \boldsymbol{\psi}_0\right) = - \overline{\boldsymbol{H}}_n^{-1}\sqrt{n} \, \boldsymbol{g}_n(\boldsymbol{\psi}_0) .
\end{equation}
Moreover, 
\begin{equation} \label{stack2}
\sqrt{n}\left(\widehat{\boldsymbol{\beta}}_{FE} - \boldsymbol{\beta}_0\right) = \left(\frac{1}{n}\sum_i \dot{\boldsymbol{X}}_i'\boldsymbol{Q}\dot{\boldsymbol{X}}_i\right)^{-1}\frac{1}{\sqrt{n}}\sum_i \dot{\boldsymbol{X}}'_i\boldsymbol{Q}\left(\dot{\boldsymbol{y}}_i - \dot{\boldsymbol{X}}_i\boldsymbol{\beta}_0 \right) 
\end{equation}

From (\ref{stack1}) and (\ref{stack2}) we see that stacking $\sqrt{n}\left(\widehat{\boldsymbol{\psi}} - \boldsymbol{\psi}_0\right)$ on top of $\sqrt{n}\left(\widehat{\boldsymbol{\beta}}_{FE} - \boldsymbol{\beta}_0\right) $ gives
\begin{equation} \label{stack}
\begin{split}
\sqrt{n}
\left(
\begin{array}{c}
  \widehat{\boldsymbol{\psi}} - \boldsymbol{\psi}_0   \\
  \widehat{\boldsymbol{\beta}}_{FE} - \boldsymbol{\beta}_0
 \end{array}
\right)
& = 
\left(
\begin{array}{cc}
\overline{\boldsymbol{H}}_n  &   \boldsymbol{0}   \\
\boldsymbol{0}  &     (1/n)\sum_i \dot{\boldsymbol{X}}_i'\boldsymbol{Q}\dot{\boldsymbol{X}}_i 
\end{array}
\right)^{-1}
\\
& \times
\left(
\begin{array}{c}
 - \sqrt{n}\,\boldsymbol{g}_n(\boldsymbol{\psi}_0)  \\
 (1/\sqrt{n})\sum_i \dot{\boldsymbol{X}}_i'\boldsymbol{Q} (\dot{\boldsymbol{y}}_i - \dot{\boldsymbol{X}}_i\boldsymbol{\beta}_0 )
\end{array}
\right)
\end{split}
\end{equation}

In order to evaluate the right-hand side of (\ref{stack}), we first show that
\begin{equation} \label{R+u+}
\left(
\begin{array}{c}
 - \sqrt{n}\,\boldsymbol{g}_n(\boldsymbol{\psi}_0)  \\
 (1/\sqrt{n})\sum_i \dot{\boldsymbol{X}}_i'\boldsymbol{Q} (\dot{\boldsymbol{y}}_i - \dot{\boldsymbol{X}}_i\boldsymbol{\beta}_0 )
\end{array}
\right)
 = \frac{1}{\sqrt{n}}\sum_i \boldsymbol{R}_i^{+ \prime}\boldsymbol{u}_i^{+ } + o_p(1).
\end{equation}
To that end, first note that in the proof of Theorem \ref{normality_thm} we found  
\begin{equation} \label{R+u+1}
\sqrt{n}\,\boldsymbol{g}_n(\boldsymbol{\psi}_0) = - \frac{1}{\sqrt{n}}\sum_{i} \boldsymbol{R}_i^{\ast \prime}\boldsymbol{u}_i + o_p(1).
\end{equation}
It remains to verify 
\begin{equation} \label{R+u+2}
\frac{1}{\sqrt{n}} \sum_i \dot{\boldsymbol{X}}_i'\boldsymbol{Q} (\dot{\boldsymbol{y}}_i - \dot{\boldsymbol{X}}_i\boldsymbol{\beta}_0 ) = \frac{1}{\sqrt{n}} \sum_i \boldsymbol{X}_i^{\ast \prime}\boldsymbol{Q} \left(\boldsymbol{y}_i^{\ast} - \boldsymbol{X}_i^{\ast}\boldsymbol{\beta}_0 \right) + o_p(1)
\end{equation}
In order to establish (\ref{R+u+2}), first observe that if we let $\boldsymbol{X}_i^{\ast} := \boldsymbol{X}_i - E(\boldsymbol{X}_1)$ and $\overline{\boldsymbol{X}}^{\ast} := (1/n)\sum_i \boldsymbol{X}_i^{\ast}$, then $
(1/\sqrt{n})\sum_i \dot{\boldsymbol{X}}_i'\boldsymbol{Q} (\dot{\boldsymbol{y}}_i - \dot{\boldsymbol{X}}_i\boldsymbol{\beta}_0 )   = (1/\sqrt{n})\sum_i \boldsymbol{X}_i^{\ast \prime}\boldsymbol{Q} (\dot{\boldsymbol{y}}_i - \dot{\boldsymbol{X}}_i\boldsymbol{\beta}_0 ) 
 - (1/\sqrt{n})\sum_i \overline{\boldsymbol{X}}^{\ast \prime}\boldsymbol{Q}(\dot{\boldsymbol{y}}_i - \dot{\boldsymbol{X}}_i\boldsymbol{\beta}_0 )
 $.  But
$\sum_i \overline{\boldsymbol{X}}^{\ast \prime}\boldsymbol{Q}(\dot{\boldsymbol{y}}_i - \dot{\boldsymbol{X}}_i\boldsymbol{\beta}_0) =  \overline{\boldsymbol{X}}^{\ast \prime}\boldsymbol{Q}\sum_i(\dot{\boldsymbol{y}}_i - \dot{\boldsymbol{X}}_i\boldsymbol{\beta}_0) = \boldsymbol{0}$.  Hence, $(1/\sqrt{n})\sum_i \dot{\boldsymbol{X}}_i'\boldsymbol{Q} (\dot{\boldsymbol{y}}_i - \dot{\boldsymbol{X}}_i\boldsymbol{\beta}_0 ) = (1/\sqrt{n})\sum_i \boldsymbol{X}_i^{\ast \prime} \boldsymbol{Q} (\dot{\boldsymbol{y}}_i - \dot{\boldsymbol{X}}_i\boldsymbol{\beta}_0 ) $. Moreover, upon setting $\boldsymbol{y}_i^{\ast} := \boldsymbol{y}_i - E(\boldsymbol{y}_1)$ and $\overline{\boldsymbol{y}}^{\ast} := (1/n)\sum_i \boldsymbol{y}_i^{\ast}$, we get $\dot{\boldsymbol{y}}_i - \dot{\boldsymbol{X}}_i\boldsymbol{\beta}_0 = \boldsymbol{y}_i^{\ast} - \overline{\boldsymbol{y}}^{\ast} - (\boldsymbol{X}_i^{\ast} - \overline{\boldsymbol{X}}^{\ast})\boldsymbol{\beta}_0 = \boldsymbol{y}_i^{\ast} - \boldsymbol{X}_i^{\ast}\boldsymbol{\beta}_0 - ( \overline{\boldsymbol{y}}^{\ast}- \overline{\boldsymbol{X}}^{\ast}\boldsymbol{\beta}_0)$. And $(1/\sqrt{n})\sum_i \boldsymbol{X}_i^{\ast \prime}\boldsymbol{Q}( \overline{\boldsymbol{y}}^{\ast}- \overline{\boldsymbol{X}}^{\ast}\boldsymbol{\beta}_0) = (1/n)\sum_i \boldsymbol{X}_i^{\ast \prime}\boldsymbol{Q}\sqrt{n}( \overline{\boldsymbol{y}}^{\ast}- \overline{\boldsymbol{X}}^{\ast}\boldsymbol{\beta}_0)$. Furthermore, $E(\sqrt{n}\overline{\boldsymbol{y}}^{\ast}) = \boldsymbol{0}$ and $\text{Var}(\sqrt{n}\overline{\boldsymbol{y}}^{\ast}) = \text{Var}(\boldsymbol{y}_1)$, which implies $\sqrt{n}\overline{\boldsymbol{y}}^{\ast} = O_p(1)$.  Similarly, $\sqrt{n}\overline{\boldsymbol{X}}^{\ast} = O_p(1)$.  On the other hand, $ (1/n)\sum_i \boldsymbol{X}_i^{\ast \prime}\boldsymbol{Q} \rightarrow_p \boldsymbol{0}$.  These observations verify (\ref{R+u+2}).  And (\ref{R+u+2}) and (\ref{R+u+1}) establish (\ref{R+u+}).

Therefore, we can evaluate the asymptotic distribution of the left-hand side of (\ref{R+u+}) by evaluating the asymptotic distribution of $(1/\sqrt{n})\sum_i \boldsymbol{R}_i^{+ \prime}\boldsymbol{u}_i^{+ }$.
To that end, consider that $E(\boldsymbol{R}_1^{\ast}\boldsymbol{u}_1) = \boldsymbol{0}$ regardless of the validity of the null hypothesis.  Moreover, suppose 
$\boldsymbol{\Lambda}_0 = \boldsymbol{\iota}_T$.  Then, because $\boldsymbol{Q}\boldsymbol{\iota}_T = \boldsymbol{0}$, we have  $E[\boldsymbol{X}_1^{\ast \prime}\boldsymbol{Q} \left(\boldsymbol{y}_1^{\ast} - \boldsymbol{X}_1^{\ast}\boldsymbol{\beta}_0 \right)] = E[\boldsymbol{X}_1^{\ast \prime}\boldsymbol{Q} (\boldsymbol{\iota}_T\boldsymbol{Z}_1^{\ast}\boldsymbol{\theta}_0 + \boldsymbol{u}_1 )] = E(\boldsymbol{X}_1^{\ast \prime}\boldsymbol{Q}  \boldsymbol{u}_1 ) = \boldsymbol{0}$.  Suppose, on the other hand, instead of  $\boldsymbol{\Lambda}_0 = \boldsymbol{\iota}_T$, we have $\boldsymbol{\Phi}_{0t}^{x} = \boldsymbol{\Phi}_{0}^{x}$ for all $t$.   From (\ref{model1}), we see that $\boldsymbol{y}_i^{\ast} - \boldsymbol{X}_i^{\ast}\boldsymbol{\beta}_0  = \boldsymbol{\Phi}_0\widetilde{\boldsymbol{\eta}}_i + \boldsymbol{\epsilon}_i$. Thus, $E[\boldsymbol{X}_1^{\ast \prime}\boldsymbol{Q} \left(\boldsymbol{y}_1^{\ast} - \boldsymbol{X}_1^{\ast}\boldsymbol{\beta}_0 \right)] = E[\boldsymbol{X}_1^{\ast \prime}\boldsymbol{Q}(\boldsymbol{\Phi}_0\widetilde{\boldsymbol{\eta}}_1 + \boldsymbol{\epsilon}_1 )] = E(\boldsymbol{X}_1^{\ast \prime}\boldsymbol{Q}\boldsymbol{\Phi}_0\widetilde{\boldsymbol{\eta}}_1)$, where the last equality follows from the assumed exogeneity of the explanatory variables. It follows from (\ref{x_ie}), and the assumption  $\boldsymbol{\Phi}_{0t}^{x} = \boldsymbol{\Phi}_{0}^{x}$ for all $t$, that $\boldsymbol{X}_i^{\ast} = \boldsymbol{\iota}_T \otimes \widetilde{\boldsymbol{\eta}}_i'\boldsymbol{\Phi}_0^{x} + \boldsymbol{E}_i^{x}$, where $\boldsymbol{E}_i^{x}$ is a $T \times K$ matrix with  $\boldsymbol{\epsilon}_{it}^{x \prime}$ in its $t$th row.  Thus, $\boldsymbol{X}_i^{\ast \prime}\boldsymbol{Q} = (\boldsymbol{\iota}_T' \otimes \boldsymbol{\Phi}_0^{x \prime}\widetilde{\boldsymbol{\eta}}_i)(\boldsymbol{Q} \otimes 1) + \boldsymbol{E}_i^{x \prime}\boldsymbol{Q} = \boldsymbol{\iota}_T' \boldsymbol{Q} \otimes \boldsymbol{\Phi}_0^{x \prime}\widetilde{\boldsymbol{\eta}}_i + \boldsymbol{E}_i^{x \prime}\boldsymbol{Q} = \boldsymbol{E}_i^{x \prime}\boldsymbol{Q}$.  Therefore, if $\boldsymbol{\Phi}_{0t}^{x} = \boldsymbol{\Phi}_{0}^{x}$ for all $t$, then $E[\boldsymbol{X}_1^{\ast \prime}\boldsymbol{Q} \left(\boldsymbol{y}_1^{\ast} - \boldsymbol{X}_1^{\ast}\boldsymbol{\beta}_0 \right)] = E(\boldsymbol{E}_1^{x \prime}\boldsymbol{Q}\boldsymbol{\Phi}_0\widetilde{\boldsymbol{\eta}}_1) = \boldsymbol{0}$, where the second equality follows from the fact that all of the entries in $\boldsymbol{E}_i^{x}$ are uncorrelated with all of the entries in $\widetilde{\boldsymbol{\eta}}_i$ by construction. The preceding shows that if $\boldsymbol{\Lambda}_0 = \boldsymbol{\iota}_T$, or $\boldsymbol{\Phi}_{0t}^{x} = \boldsymbol{\Phi}_{0}^{x}$ for all $t$, or both, then $E(\boldsymbol{R}_1^{+ \prime}\boldsymbol{u}_1^{+ }) = \boldsymbol{0}$.  Moreover, Assumption A2$^{\ast}$ implies $\text{Var}(\boldsymbol{R}_1^{+ \prime}\boldsymbol{u}_1^{+ }) = E(\boldsymbol{R}_1^{+ \prime}\boldsymbol{u}_1^{+ }\boldsymbol{u}_1^{+ \prime}\boldsymbol{R}_1^{+}) = \boldsymbol{A}_0^+$ is defined. It follows from the central limit theorem for i.i.d. random vectors that $(1/\sqrt{n})\sum_i \boldsymbol{R}_i^{+ \prime}\boldsymbol{u}_i^{+ } \rightarrow_d N(\boldsymbol{0},\boldsymbol{A}_0^+)$ \citep[see, .e.g.,][p. 21, Theorem 1.2.4]{Greenberg1983}.

It remains to evaluate the large sample behavior of the first matrix on the right-hand side of (\ref{stack}).  

Note that
\begin{equation*}
\begin{split}
 \frac{1}{n}\sum_i \dot{\boldsymbol{X}}_i'\boldsymbol{Q}\dot{\boldsymbol{X}}_i   &=  \frac{1}{n}\sum_i \left(\boldsymbol{X}_i^{\ast} - \overline{\boldsymbol{X}}^{\ast} \right)'\boldsymbol{Q}\left(\boldsymbol{X}_i^{\ast} - \overline{\boldsymbol{X}}^{\ast} \right)  \\
 &=  \frac{1}{n}\sum_i \boldsymbol{X}_i^{\ast \prime}\boldsymbol{Q}\boldsymbol{X}_i^{\ast}  -   \overline{\boldsymbol{X}}^{\ast \prime}\boldsymbol{Q} \overline{\boldsymbol{X}}^{\ast} ,
 \end{split}
\end{equation*}
and $\overline{\boldsymbol{X}}^{\ast} \rightarrow_p \boldsymbol{0}$.  Hence, $(1/n)\sum_i \dot{\boldsymbol{X}}_i'\boldsymbol{Q}\dot{\boldsymbol{X}}_i   =  (1/n)\sum_i \boldsymbol{X}_i^{\ast \prime}\boldsymbol{Q}\boldsymbol{X}_i^{\ast} + o_p(1)$. Moreover, by the law of large numbers, $(1/n)\sum_i \boldsymbol{X}_i^{\ast \prime}\boldsymbol{Q}\boldsymbol{X}_i^{\ast} \rightarrow_p E(\boldsymbol{X}_1^{\ast \prime}\boldsymbol{Q}\boldsymbol{X}_1^{\ast})$. Hence, $(1/n)\sum_i \dot{\boldsymbol{X}}_i'\boldsymbol{Q}\dot{\boldsymbol{X}}_i \rightarrow_p E(\boldsymbol{X}_1^{\ast \prime}\boldsymbol{Q}\boldsymbol{X}_1^{\ast})$.

Finally, we show that $\overline{\boldsymbol{H}}_n \rightarrow_p \boldsymbol{H}_0$.  To that end, note that the ($j,k$)th entry in $\overline{\boldsymbol{H}}_n$ is $(1/n)\sum_i \partial^2 s(\dot{\boldsymbol{v}}_i,\overline{\boldsymbol{\psi}}_{(j)})/\partial{\psi}_j\partial{\psi}_k$. Moreover, if  $q=1$, then $\boldsymbol{\Lambda}' = (1, \boldsymbol{\lambda}') = (1,\lambda_2,\ldots,\lambda_T)$ and $\boldsymbol{\theta}'= (\theta_1,\ldots,\theta_m)$.  It follows that
\begin{align*}
	\frac{\partial^2 s(\dot{\boldsymbol{v}}_i,\boldsymbol{\psi})}{\partial \boldsymbol{\gamma} \partial \boldsymbol{\gamma}^{\prime}} & = 
\left(
\begin{array}{cc}
  \dot{\boldsymbol{X}}_i^{ \prime}\dot{\boldsymbol{X}}_i & \left(\dot{\boldsymbol{x}}_{i1} + \dot{\boldsymbol{X}}_{2i}'\boldsymbol{\lambda} \right)\dot{\boldsymbol{z}}_i'   \\
 \dot{\boldsymbol{z}}_i\left(\dot{\boldsymbol{x}}_{i1}' + \boldsymbol{\lambda}'\dot{\boldsymbol{X}}_{2i} \right)   &      \left(1 + \boldsymbol{\lambda}'\boldsymbol{\lambda} \right)\dot{\boldsymbol{z}}_i\dot{\boldsymbol{z}}_i' 
\end{array}
\right),
 \\
	\frac{\partial^2 s(\dot{\boldsymbol{v}}_i,\boldsymbol{\psi})}{\partial \boldsymbol{\lambda} \partial \boldsymbol{\beta}^{\prime}} & =  \left( \dot{\boldsymbol{z}}_i' \boldsymbol{\theta} \otimes \boldsymbol{I}_{T-1} \right)  \dot{\boldsymbol{X}}_{2i} =  \dot{\boldsymbol{z}}_i' \boldsymbol{\theta}   \dot{\boldsymbol{X}}_{2i}, \\
	\frac{\partial^2 s(\dot{\boldsymbol{v}}_i,\boldsymbol{\psi})}{\partial \boldsymbol{\lambda} \partial \boldsymbol{\theta}^{\prime}} & = - \left( \dot{\boldsymbol{y}}_{2i} - \dot{\boldsymbol{X}}_{2i}\boldsymbol{\beta} - \boldsymbol{\lambda}\dot{\boldsymbol{z}}_i'\boldsymbol{\theta} \right) \dot{\boldsymbol{z}}_i^{ \prime} + \dot{\boldsymbol{z}}_i^{ \prime} \boldsymbol{\theta} \boldsymbol{\lambda}\dot{\boldsymbol{z}}_i^{ \prime} \\
	\frac{\partial^2 s(\dot{\boldsymbol{v}}_i,\boldsymbol{\psi})}{\partial \boldsymbol{\lambda} \partial \boldsymbol{\lambda}^{\prime}} & = \left(\dot{\boldsymbol{z}}_i^{ \prime} \boldsymbol{\theta}\right)^2 \boldsymbol{I}_{T-q}.
\end{align*}

Define $\widetilde{\boldsymbol{\psi}}$ so that $\widetilde{\boldsymbol{\psi}} \rightarrow_p \boldsymbol{\psi}_0$, and note that
\begin{equation*}
\begin{split}
\frac{1}{n}\sum_i \frac{\partial^2 s(\dot{\boldsymbol{v}}_i,\widetilde{\boldsymbol{\psi}})}{\partial \boldsymbol{\gamma} \partial \boldsymbol{\gamma}^{\prime}}  & - \frac{1}{n}\sum_i \frac{\partial^2 s(\dot{\boldsymbol{v}}_i,\boldsymbol{\psi}_0)}{\partial \boldsymbol{\gamma} \partial \boldsymbol{\gamma}^{\prime}}  \\
& = 
\left(
\begin{array}{cc}
  \boldsymbol{0} &  (1/n)\sum_i \dot{\boldsymbol{X}}_{2i}'(\widetilde{\boldsymbol{\lambda}} - \boldsymbol{\lambda}_0 )\dot{\boldsymbol{z}}_i'   \\
 (1/n)\sum_i \dot{\boldsymbol{z}}_i(\widetilde{\boldsymbol{\lambda}} - \boldsymbol{\lambda}_0 )' \dot{\boldsymbol{X}}_{2i}   &      (\widetilde{\boldsymbol{\lambda}}'\widetilde{\boldsymbol{\lambda}} - \boldsymbol{\lambda}'_0\boldsymbol{\lambda} _0)(1/n)\sum_i\dot{\boldsymbol{z}}_i\dot{\boldsymbol{z}}_i' 
\end{array} 
\right).
\end{split}
\end{equation*}
Our assumptions ensure $(1/n)\sum_i\dot{\boldsymbol{z}}_i\dot{\boldsymbol{z}}_i' = O_p(1)$.  Moreover, $\widetilde{\boldsymbol{\lambda}}'\widetilde{\boldsymbol{\lambda}} - \boldsymbol{\lambda}'_0\boldsymbol{\lambda} _0 \rightarrow_p 0$.  Hence, $(\widetilde{\boldsymbol{\lambda}}'\widetilde{\boldsymbol{\lambda}} - \boldsymbol{\lambda}'_0\boldsymbol{\lambda} _0)(1/n)\sum_i\dot{\boldsymbol{z}}_i\dot{\boldsymbol{z}}_i' \rightarrow_p \boldsymbol{0}$.  Also, $(1/n)\sum_i \dot{\boldsymbol{X}}_{2i}'(\widetilde{\boldsymbol{\lambda}} - \boldsymbol{\lambda}_0 )\dot{\boldsymbol{z}}_i' = (1/n)\sum_i \sum_{t=2}^T\dot{\boldsymbol{x}}_{it}(\widetilde{\lambda}_t - \lambda_{0t} )\dot{\boldsymbol{z}}_i' $. The $(j,k)$th entry in the latter $(T-1) \times m$ matrix is $(1/n)\sum_i \sum_{t=2}^T\dot{x}_{itj}\dot{z}_{ik}(\widetilde{\lambda}_t - \lambda_{0t} )$, where $\dot{x}_{itj}$ is the $j$th entry in $\dot{\boldsymbol{x}}_{it}$ and $\dot{z}_{ik}$ is the $k$th entry in $\dot{\boldsymbol{z}}_{i}$. Next, note that 
\begin{equation*}
\begin{split}
\left| \frac{1}{n}\sum_i \sum_{t=2}^T\dot{x}_{itk}\dot{z}_{ik}(\widetilde{\lambda}_t - \lambda_{0t} )\right|  & \le    \sum_{t=2}^T\left|(\widetilde{\lambda}_t - \lambda_{0t} )\frac{1}{n}\sum_i\dot{x}_{itk}\dot{z}_{ik}\right| \\
& \le \max_{t}\left| \widetilde{\lambda}_t - \lambda_{0t} \right| \sum_{t=2}^T\left| \frac{1}{n}\sum_i\dot{x}_{itk}\dot{z}_{ik}\right|.
\end{split}
\end{equation*}
 Elsewhere we have shown that $(1/n)\sum_i\dot{x}_{itk}\dot{z}_{ik} \rightarrow_p E(x_{1tk}^{\ast}z_{ik}^{\ast})$.  And $\max_{t}\left| \widetilde{\lambda}_t - \lambda_{0t} \right| \rightarrow_p 0$.  The preceding shows that $\partial^2 S_n(\widetilde{\boldsymbol{\psi}})/\partial \boldsymbol{\gamma} \partial \boldsymbol{\gamma}' - \partial^2 S_n(\boldsymbol{\psi}_0)/\partial \boldsymbol{\gamma} \partial \boldsymbol{\gamma}'$ for any $\widetilde{\boldsymbol{\psi}}$ such that $\widetilde{\boldsymbol{\psi}} \rightarrow_p \boldsymbol{\psi}_0$.  Similarly arguments can be used to show that $\partial^2 S_n(\widetilde{\boldsymbol{\psi}})/\partial \psi_j \partial \psi_k - \partial^2 S_n(\boldsymbol{\psi}_0)/\partial \psi_j \partial \psi_k$ for all $j$, $k$,  and $\widetilde{\boldsymbol{\psi}}$ such that $\widetilde{\boldsymbol{\psi}} \rightarrow_p \boldsymbol{\psi}_0$.  Hence, $\partial^2 S_n(\overline{\boldsymbol{\psi}}_{(j)})/\partial \psi_j \partial \psi_k - \partial^2 S_n(\boldsymbol{\psi}_0)/\partial \psi_j \partial \psi_k \rightarrow_p 0$, because $\parallel \overline{\boldsymbol{\psi}}_{(j)} - \boldsymbol{\psi}_0 \parallel \le \parallel \widehat{\boldsymbol{\psi}} - \boldsymbol{\psi}_0 \parallel$ and our assumptions imply $\widehat{\boldsymbol{\psi}} \rightarrow_p \boldsymbol{\psi}_0$, and thus $\overline{\boldsymbol{\psi}}_{(j)} \rightarrow_p \boldsymbol{\psi}_0$ ($j=1,\ldots,p$). In other words, the preceding shows that $\overline{\boldsymbol{H}}_n - \boldsymbol{H}_n(\boldsymbol{\psi}_0) \rightarrow_p \boldsymbol{0}$.  Moreover, in the proof of Theorem \ref{normality_thm} we showed that  $\boldsymbol{H}_n(\boldsymbol{\psi}_0) \rightarrow_p \boldsymbol{H}_0 $.  Hence, $\overline{\boldsymbol{H}}_n \rightarrow_p \boldsymbol{H}_0$.

The preceding verifies that the first matrix on the right-hand side of (\ref{stack}) converges in probability to $(\boldsymbol{H}_0^+)^{-1}$ while the second matrix  converges in distribution to a multivariate normal distribution with mean $\boldsymbol{0}$ and variance-covariance matrix $\boldsymbol{A}_0^+$, provided $\boldsymbol{\Lambda}_0 = \boldsymbol{\iota}_T$, or $\boldsymbol{\Phi}_{0t}^{x} = \boldsymbol{\Phi}_{0}^{x}$ for all $t$, or both.  Hence, if $\boldsymbol{\Lambda}_0 = \boldsymbol{\iota}_T$, or $\boldsymbol{\Phi}_{0t}^{x} = \boldsymbol{\Phi}_{0}^{x}$ for all $t$, or both, then
\begin{equation*}
\sqrt{n}
\left(
\begin{array}{c}
  \widehat{\boldsymbol{\psi}} - \boldsymbol{\psi}_0   \\
  \widehat{\boldsymbol{\beta}}_{FE} - \boldsymbol{\beta}_0
 \end{array}
\right)
\rightarrow_d N(\,\boldsymbol{0},\,(\boldsymbol{H}_0^+)^{-1}\boldsymbol{A}_0^+(\boldsymbol{H}_0^+)^{-1}\,).
\end{equation*}

\subsection*{A.4 $\;$ Theorem \ref{computation_thm} proof}

For any given $\boldsymbol{\lambda}$, the objective function in (\ref{FWL_obj_fcn}) is minimized with respect to $\boldsymbol{\beta}$ by
\begin{equation*} 
\widehat{\boldsymbol{\beta}}(\boldsymbol{\lambda}) := \left(\dot{\boldsymbol{X}}'\boldsymbol{M}(\boldsymbol{\lambda})\dot{\boldsymbol{X}}\right)^{-1} \dot{\boldsymbol{X}}'\boldsymbol{M} (\boldsymbol{\lambda})\dot{\boldsymbol{y}}.
\end{equation*}
Hence, if $\widetilde{\boldsymbol{\alpha}} = (\widetilde{\boldsymbol{\beta}}' ,\widetilde{\boldsymbol{\lambda}}')'$ minimizes the objective function in (\ref{FWL_obj_fcn}), it must be that
\begin{equation} \label{FWL_beta}
\widetilde{\boldsymbol{\beta}} =  \left(\dot{\boldsymbol{X}}'\boldsymbol{M}(\widetilde{\boldsymbol{\lambda}})\dot{\boldsymbol{X}}\right)^{-1} \dot{\boldsymbol{X}}'\boldsymbol{M} (\widetilde{\boldsymbol{\lambda}})\dot{\boldsymbol{y}}.
\end{equation}

We next show that the formula in (\ref{FWL_beta}) is equivalent to the formula in Equation (\ref{PIE_soln}) in Theorem \ref{computation_thm}.
To that end, observe that there is a the  $mT\times K$ matrix  of zeros and ones, say $\boldsymbol{S}$, such that $\dot{\boldsymbol{X}}=(\dot{\boldsymbol{z}}\otimes \boldsymbol{I}_T)\boldsymbol{S}$, in which case,  $\boldsymbol{M}(\boldsymbol{\lambda})\dot{\boldsymbol{X}}=\boldsymbol{M}(\boldsymbol{\lambda})(\dot{\boldsymbol{z}}\otimes \boldsymbol{I}_T)\boldsymbol{S}= \dot{\boldsymbol{X}}-\left(\dot{\boldsymbol{z}} \otimes \boldsymbol{P}_{\boldsymbol{\Lambda}} \right)\boldsymbol{S}= \dot{\boldsymbol{X}}-\left(\boldsymbol{I}_n \otimes \boldsymbol{P}_{\boldsymbol{\Lambda}} \right)(\dot{\boldsymbol{z}}\otimes \boldsymbol{I}_T)\boldsymbol{S}=\boldsymbol{M}^{\ast}(\boldsymbol{\lambda})\dot{\boldsymbol{X}}$, where $
	\boldsymbol{M}^{\ast}(\boldsymbol{\lambda})=\boldsymbol{I}_{nT}-\left(\boldsymbol{I}_n \otimes \boldsymbol{P}_{\boldsymbol{\Lambda}} \right)$.
Therefore, 
\begin{equation*}
\begin{split}
	\widehat{\boldsymbol{\beta}}(\boldsymbol{\lambda}) & = \left(\dot{\boldsymbol{X}}'\boldsymbol{M}(\boldsymbol{\lambda})\dot{\boldsymbol{X}}\right)^{-1} \dot{\boldsymbol{X}}'\boldsymbol{M}(\boldsymbol{\lambda})\dot{\boldsymbol{y}} \\
	& = \left(\dot{\boldsymbol{X}}'\boldsymbol{M}^{\ast}(\boldsymbol{\lambda})\dot{\boldsymbol{X}}\right)^{-1} \dot{\boldsymbol{X}}'\boldsymbol{M}^{\ast} (\boldsymbol{\lambda})\dot{\boldsymbol{y}} \\
	& =  \left( \sum_i \dot{\boldsymbol{X}}_i'\boldsymbol{Q}(\boldsymbol{\lambda}) \dot{\boldsymbol{X}}_i\right) ^{-1} \sum_i \dot{\boldsymbol{X}}_i'\boldsymbol{Q}(\boldsymbol{\lambda}) \dot{\boldsymbol{y}}_i.
\end{split}
\end{equation*}
This result and Equation (\ref{FWL_beta}) give Equation (\ref{PIE_soln}).

It remains to verify Eq. (\ref{min_S_n}).  To do that, first note that $\dot{\boldsymbol{Z}}_i\boldsymbol{\theta}=(I_q\otimes \dot{\boldsymbol{z}}_i')\text{vec}(\boldsymbol{\Theta})$ by definition.  Moreover, using a relationship between vectorization and Kronecker product, we have $(I_q\otimes \dot{\boldsymbol{z}}_i')\text{vec}(\boldsymbol{\Theta})=\text{vec}(\dot{\boldsymbol{z}}_i' \boldsymbol{\Theta})$.  Moreover, calculations show that $\text{vec}(\dot{\boldsymbol{z}}_i' \boldsymbol{\Theta}) = \text{vec}(\boldsymbol{\Theta}'\dot{\boldsymbol{z}}_i)$. And again, on exploiting a relationship between vectorization and Kronecker product, we have $\text{vec}(\boldsymbol{\Theta}'\dot{\boldsymbol{z}}_i) = (\dot{\boldsymbol{z}}_i' \otimes \boldsymbol{I}_q)\text{vec}(\boldsymbol{\Theta}')$.  Hence, $ \boldsymbol{\Lambda} \dot{\boldsymbol{Z}}_i\boldsymbol{\theta} = \boldsymbol{\Lambda}(\dot{\boldsymbol{z}}_i' \otimes \boldsymbol{I}_q)\text{vec}(\boldsymbol{\Theta}')  = (1 \otimes \boldsymbol{\Lambda})(\dot{\boldsymbol{z}}_i' \otimes \boldsymbol{I}_q)\text{vec}(\boldsymbol{\Theta}')= (\dot{\boldsymbol{z}}_i' \otimes \boldsymbol{\Lambda})\text{vec}(\boldsymbol{\Theta}') = (\dot{\boldsymbol{z}}_i' \otimes \boldsymbol{\Lambda})\boldsymbol{\theta}^o$, where $ \boldsymbol{\theta}^{o} := \text{vec}(\boldsymbol{\Theta}')$.  Stacking the $T\times 1$ vectors $(\dot{\boldsymbol{z}}_i' \otimes \boldsymbol{\Lambda})\boldsymbol{\theta}^o$ ($i=1,\ldots,n$) on top of each other we get the $nT\times 1$ vector $(\dot{\boldsymbol{z}} \otimes \boldsymbol{\Lambda})\boldsymbol{\theta}^o$.  Moreover, let $\dot{\boldsymbol{Z}} := (\dot{\boldsymbol{Z}}_1',\ldots, \dot{\boldsymbol{Z}}_n')'$. Then, it follows from the preceding that 
\begin{equation} \label{S_n}
\begin{split}
S_n(\boldsymbol{\psi}) &= \frac{1}{2n} \left( \dot{\boldsymbol{y}} -\dot{\boldsymbol{X}}\boldsymbol{\beta} - (\boldsymbol{I}_n \otimes \boldsymbol{\Lambda}) \dot{\boldsymbol{Z}}\boldsymbol{\theta}\right)'\left( \dot{\boldsymbol{y}} -\dot{\boldsymbol{X}}\boldsymbol{\beta} - (\boldsymbol{I}_n \otimes \boldsymbol{\Lambda}) \dot{\boldsymbol{Z}}\boldsymbol{\theta}\right) \\
 &=   \frac{1}{2n} \left( \dot{\boldsymbol{y}} -\dot{\boldsymbol{X}}\boldsymbol{\beta} - (\dot{\boldsymbol{z}} \otimes \boldsymbol{\Lambda}) \boldsymbol{\theta}^{o}\right)'\left( \dot{\boldsymbol{y}} -\dot{\boldsymbol{X}}\boldsymbol{\beta} - (\dot{\boldsymbol{z}} \otimes \boldsymbol{\Lambda}) \boldsymbol{\theta}^{o}\right).
\end{split}
\end{equation}

For any given $\boldsymbol{\lambda}$, the sum of squares
\begin{equation*}
   \left( \dot{\boldsymbol{y}} -\dot{\boldsymbol{X}}\boldsymbol{\beta} - (\dot{\boldsymbol{z}} \otimes \boldsymbol{\Lambda}) \boldsymbol{\theta}^{o}\right)'\left( \dot{\boldsymbol{y}} -\dot{\boldsymbol{X}}\boldsymbol{\beta} - (\dot{\boldsymbol{z}} \otimes \boldsymbol{\Lambda}) \boldsymbol{\theta}^{o}\right).
\end{equation*}
is minimized by the least-squares estimates of $\boldsymbol{\beta}$ and $\boldsymbol{\theta}^o$.  Moreover, according the FWL theorem, these estimates are given by
\begin{equation} \label{theta_given_lambda}
	\widehat{\boldsymbol{\theta}}^o(\boldsymbol{\lambda}) := \left( (\dot{\boldsymbol{z}} \otimes \boldsymbol{\Lambda})'\boldsymbol{M}(\dot{\boldsymbol{z}} \otimes \boldsymbol{\Lambda})\right)^{-1}(\dot{\boldsymbol{z}} \otimes \boldsymbol{\Lambda})'\boldsymbol{M}\dot{\boldsymbol{y}}, 
\end{equation}
and 
\begin{equation} \label{beta_given_lambda}
	\widehat{\boldsymbol{\beta}}(\boldsymbol{\lambda}) := \left(\dot{\boldsymbol{X}}'\boldsymbol{M}(\boldsymbol{\lambda})\dot{\boldsymbol{X}}\right)^{-1} \dot{\boldsymbol{X}}'\boldsymbol{M} (\boldsymbol{\lambda})\dot{\boldsymbol{y}},
\end{equation}
where
\begin{align*}
	\boldsymbol{M}(\boldsymbol{\lambda}) &=\boldsymbol{I}_{nT}-(\dot{\boldsymbol{z}} \otimes \boldsymbol{\Lambda})
	\left((\dot{\boldsymbol{z}} \otimes \boldsymbol{\Lambda})'(\dot{\boldsymbol{z}} \otimes \boldsymbol{\Lambda})\right)^{-1}(\dot{\boldsymbol{z}} \otimes \boldsymbol{\Lambda})'\\
	&=\boldsymbol{I}_{nT}-\left(\boldsymbol{P}_{\dot{\boldsymbol{z}} } \otimes \boldsymbol{P}_{\boldsymbol{\Lambda}} \right).
\end{align*}
It follows from the preceding observations that, for each $\boldsymbol{\lambda}$, we have
\begin{equation*}
\begin{split}
  \left( \dot{\boldsymbol{y}} -\dot{\boldsymbol{X}}\widehat{\boldsymbol{\beta}}(\boldsymbol{\lambda})- (\dot{\boldsymbol{z}} \otimes \boldsymbol{\Lambda}) \widehat{\boldsymbol{\theta}}^o(\boldsymbol{\lambda})\right)' & \left( \dot{\boldsymbol{y}} -\dot{\boldsymbol{X}}\widehat{\boldsymbol{\beta}}(\boldsymbol{\lambda}) - (\dot{\boldsymbol{z}} \otimes \boldsymbol{\Lambda}) \widehat{\boldsymbol{\theta}}^o(\boldsymbol{\lambda})\right) \\
& \le \left( \dot{\boldsymbol{y}} -\dot{\boldsymbol{X}}\boldsymbol{\beta} - (\dot{\boldsymbol{z}} \otimes \boldsymbol{\Lambda}) \boldsymbol{\theta}^{o}\right)' \left( \dot{\boldsymbol{y}} -\dot{\boldsymbol{X}}\boldsymbol{\beta} - (\dot{\boldsymbol{z}} \otimes \boldsymbol{\Lambda}) \boldsymbol{\theta}^{o}\right)  
 \end{split}
\end{equation*}
for all $\boldsymbol{\beta}$ and $\boldsymbol{\theta}^o$. Note that the only parameters in the sum of squares on the left-hand side of this expression are those in $\boldsymbol{\lambda}$.  And, given the inequality in the above expression, we see that if the left-hand sum of squares is minimized with respect to $\boldsymbol{\lambda}$, then the right-hand sum of squares is minimized with respect to $\boldsymbol{\lambda}$, $\boldsymbol{\theta}^o$, and $\boldsymbol{\beta}$.  Moreover, it also follows from the FWL theorem that 
\begin{equation*}
\begin{split}
  \left( \dot{\boldsymbol{y}} -\dot{\boldsymbol{X}}\widehat{\boldsymbol{\beta}}(\boldsymbol{\lambda})- (\dot{\boldsymbol{z}} \otimes \boldsymbol{\Lambda}) \widehat{\boldsymbol{\theta}}^o(\boldsymbol{\lambda})\right)' & \left( \dot{\boldsymbol{y}} -\dot{\boldsymbol{X}}\widehat{\boldsymbol{\beta}}(\boldsymbol{\lambda}) - (\dot{\boldsymbol{z}} \otimes \boldsymbol{\Lambda}) \widehat{\boldsymbol{\theta}}^o(\boldsymbol{\lambda})\right) \\
 &= \left(\dot{\boldsymbol{y}} - \dot{\boldsymbol{X}}\widehat{\boldsymbol{\beta}}(\boldsymbol{\lambda}) \right)'\boldsymbol{M}(\boldsymbol{\lambda})\left(\dot{\boldsymbol{y}} - \dot{\boldsymbol{X}}\widehat{\boldsymbol{\beta}}(\boldsymbol{\lambda}) \right),
 \end{split}
\end{equation*}
and thus whatever $\boldsymbol{\lambda}$ minimizes the left-hand side of this expression minimizes the right-hand side, and vice versa.  However, $\widetilde{\boldsymbol{\lambda}}$ minimizes the right-hand side of the last expression.

It follows from the foregoing that, if we set $\widetilde{\boldsymbol{\Lambda}}$ equal to $\boldsymbol{\Lambda}$ evaluated at $\widetilde{\boldsymbol{\lambda}}$, we set $\widetilde{\boldsymbol{\theta}}^o := \widehat{\boldsymbol{\theta}}^o(\widetilde{\boldsymbol{\lambda}})$, and recall that $\widetilde{\boldsymbol{\beta}} = \widehat{\boldsymbol{\beta}}(\widetilde{\boldsymbol{\lambda}})$, then
\begin{equation} \label{inequality}
\begin{split}
  \left( \dot{\boldsymbol{y}} -\dot{\boldsymbol{X}}\widetilde{\boldsymbol{\beta}}- (\dot{\boldsymbol{z}} \otimes \widetilde{\boldsymbol{\Lambda}}) \widetilde{\boldsymbol{\theta}}^o\right)' & \left( \dot{\boldsymbol{y}} -\dot{\boldsymbol{X}}\widetilde{\boldsymbol{\beta}} - (\dot{\boldsymbol{z}} \otimes \widetilde{\boldsymbol{\Lambda}}) \widetilde{\boldsymbol{\theta}}^o\right) \\
& \le \left( \dot{\boldsymbol{y}} -\dot{\boldsymbol{X}}\boldsymbol{\beta} - (\dot{\boldsymbol{z}} \otimes \boldsymbol{\Lambda}) \boldsymbol{\theta}^{o}\right)' \left( \dot{\boldsymbol{y}} -\dot{\boldsymbol{X}}\boldsymbol{\beta} - (\dot{\boldsymbol{z}} \otimes \boldsymbol{\Lambda}) \boldsymbol{\theta}^{o}\right)  
 \end{split}
\end{equation}
for all $\boldsymbol{\beta}$, $\boldsymbol{\theta}^o$, and $\boldsymbol{\lambda}$.

Next, note that $\widetilde{\boldsymbol{\theta}}^o =  \text{vec}(\widehat{\boldsymbol{\Theta}}^o(\widetilde{\boldsymbol{\lambda}})')$ (see Equations (\ref{FWL_theta}) and (\ref{theta_given_lambda})).  Moreover, $\widetilde{\boldsymbol{\theta}}$ is a permutation of the entries in $\widetilde{\boldsymbol{\theta}}^o$; specifically, $\widetilde{\boldsymbol{\theta}} =  \text{vec}(\widehat{\boldsymbol{\Theta}}^o(\widetilde{\boldsymbol{\lambda}}))$. From this observation, (\ref{S_n}), and (\ref{inequality}), we get
\begin{equation*}
\begin{split}
 \left( \dot{\boldsymbol{y}} -\dot{\boldsymbol{X}}\widetilde{\boldsymbol{\beta}} - (\boldsymbol{I}_n \otimes \widetilde{\boldsymbol{\Lambda}}) \dot{\boldsymbol{Z}}\widetilde{\boldsymbol{\theta}}\right)' &\left( \dot{\boldsymbol{y}} -\dot{\boldsymbol{X}}\widetilde{\boldsymbol{\beta}} - (\boldsymbol{I}_n \otimes \widetilde{\boldsymbol{\Lambda}}) \dot{\boldsymbol{Z}}\widetilde{\boldsymbol{\theta}}\right) \\
 & =  \left( \dot{\boldsymbol{y}} -\dot{\boldsymbol{X}}\widetilde{\boldsymbol{\beta}}- (\dot{\boldsymbol{z}} \otimes \widetilde{\boldsymbol{\Lambda}}) \widetilde{\boldsymbol{\theta}}^o\right)'  \left( \dot{\boldsymbol{y}} -\dot{\boldsymbol{X}}\widetilde{\boldsymbol{\beta}} - (\dot{\boldsymbol{z}} \otimes \widetilde{\boldsymbol{\Lambda}}) \widetilde{\boldsymbol{\theta}}^o\right) \\
& \le \left( \dot{\boldsymbol{y}} -\dot{\boldsymbol{X}}\boldsymbol{\beta} - (\dot{\boldsymbol{z}} \otimes \boldsymbol{\Lambda}) \boldsymbol{\theta}^{o}\right)' \left( \dot{\boldsymbol{y}} -\dot{\boldsymbol{X}}\boldsymbol{\beta} - (\dot{\boldsymbol{z}} \otimes \boldsymbol{\Lambda}) \boldsymbol{\theta}^{o}\right)  \\
&=  \left( \dot{\boldsymbol{y}} -\dot{\boldsymbol{X}}\boldsymbol{\beta} - (\boldsymbol{I}_n \otimes \boldsymbol{\Lambda}) \dot{\boldsymbol{Z}}\boldsymbol{\theta}\right)'\left( \dot{\boldsymbol{y}} -\dot{\boldsymbol{X}}\boldsymbol{\beta} - (\boldsymbol{I}_n \otimes \boldsymbol{\Lambda}) \dot{\boldsymbol{Z}}\boldsymbol{\theta}\right) 
 \end{split}
\end{equation*}
for all $\boldsymbol{\beta}$, $\boldsymbol{\theta}$, and $\boldsymbol{\lambda}$.  This verifies (\ref{min_S_n}).  Given $\widehat{\boldsymbol{\psi}}$ denotes the NLS estimate, the preceding verifies that $\widehat{\boldsymbol{\psi}}=\widetilde{\boldsymbol{\psi}}$.

\section*{Appendix B }

Recall that $ \dot{\boldsymbol{e}}(\boldsymbol{\beta}) =\dot{\boldsymbol{y}}-\dot{\boldsymbol{X}}\boldsymbol{\beta} $ and note that the objective function $\dot{\boldsymbol{e}}(\boldsymbol{\beta})'\boldsymbol{M}(\boldsymbol{\lambda})\dot{\boldsymbol{e}}(\boldsymbol{\beta})$ depends on $\boldsymbol{\Lambda}$ only through $\boldsymbol{M}(\boldsymbol{\lambda})$. Therefore, for a fixed $\boldsymbol{\beta}$, the optimal $\boldsymbol{\Lambda}$ minimizes $  \dot{\boldsymbol{e}}(\boldsymbol{\beta})'\boldsymbol{M}(\boldsymbol{\lambda})\dot{\boldsymbol{e}}(\boldsymbol{\beta})$ and hence maximizes $  \dot{\boldsymbol{e}}(\boldsymbol{\beta})'\left(\boldsymbol{P}_{\dot{\boldsymbol{z}} } \otimes \boldsymbol{P}_{\boldsymbol{\Lambda}} \right)\dot{\boldsymbol{e}}(\boldsymbol{\beta})$.  Moreover, recall that $ \boldsymbol{E}(
\boldsymbol{\beta})$ is a $ n \times T $ matrix such that  $\text{vec}\left( \boldsymbol{E}(\boldsymbol{\beta})'\right) =\dot{\boldsymbol{e}}(\boldsymbol{\beta})$.  This and known vectorization results, give $ \left(\boldsymbol{P}_{\dot{\boldsymbol{z}} } \otimes \boldsymbol{P}_{\boldsymbol{\Lambda}} \right)\dot{\boldsymbol{e}}(\boldsymbol{\beta}) = \left(\boldsymbol{P}_{\dot{\boldsymbol{z}}} \otimes \boldsymbol{P}_{\boldsymbol{\Lambda}} \right)\text{vec}\left( \boldsymbol{E}(\boldsymbol{\beta})'\right) = \text{vec}\left( \boldsymbol{P}_{\boldsymbol{\Lambda}}\boldsymbol{E}(\boldsymbol{\beta})' \boldsymbol{P}_{\dot{\boldsymbol{z}}} \right) $ Furthermore, the matrices $ \boldsymbol{P}_{\dot{\boldsymbol{z}} } $ and $ \boldsymbol{P}_{\boldsymbol{\Lambda}} $  are symmetric and idempotent. Hence, $  \dot{\boldsymbol{e}}(\boldsymbol{\beta})'\left(\boldsymbol{P}_{\dot{\boldsymbol{z}} } \otimes \boldsymbol{P}_{\boldsymbol{\Lambda}} \right)\dot{\boldsymbol{e}}(\boldsymbol{\beta}) =  \text{vec}\left( \boldsymbol{P}_{\boldsymbol{\Lambda}}\boldsymbol{E}(\boldsymbol{\beta})' \boldsymbol{P}_{\dot{\boldsymbol{z}}} \right)'\text{vec}\left( \boldsymbol{P}_{\boldsymbol{\Lambda}}\boldsymbol{E}(\boldsymbol{\beta})' \boldsymbol{P}_{\dot{\boldsymbol{z}}} \right)$.  Moreover, $ \text{vec}\left( \boldsymbol{P}_{\boldsymbol{\Lambda}}\boldsymbol{E}(\boldsymbol{\beta})' \boldsymbol{P}_{\dot{\boldsymbol{z}}} \right)'\text{vec}\left( \boldsymbol{P}_{\boldsymbol{\Lambda}}\boldsymbol{E}(\boldsymbol{\beta})' \boldsymbol{P}_{\dot{\boldsymbol{z}}} \right) = \text{tr}\left[ \left( \boldsymbol{P}_{\boldsymbol{\Lambda}}\boldsymbol{E}(\boldsymbol{\beta})' \boldsymbol{P}_{\dot{\boldsymbol{z}}} \right)'  \boldsymbol{P}_{\boldsymbol{\Lambda}}\boldsymbol{E}(\boldsymbol{\beta})' \boldsymbol{P}_{\dot{\boldsymbol{z}}} \right]  =  \text{tr} \left( \boldsymbol{P}_{\dot{\boldsymbol{z}}}\boldsymbol{E}(\boldsymbol{\beta})   \boldsymbol{P}_{\boldsymbol{\Lambda}}\boldsymbol{E}(\boldsymbol{\beta})' \boldsymbol{P}_{\dot{\boldsymbol{z}}} \right) $.  Let $\boldsymbol{\Lambda}^{\ast}=\boldsymbol{\Lambda}(\boldsymbol{\Lambda}'\boldsymbol{\Lambda})^{-1/2}$ so that $\boldsymbol{P_{\Lambda}}= \boldsymbol{\Lambda}^{\ast}\boldsymbol{\Lambda}^{\ast \prime}$. Then  $\text{tr} \left( \boldsymbol{P}_{\dot{\boldsymbol{z}}}\boldsymbol{E}(\boldsymbol{\beta})   \boldsymbol{P}_{\boldsymbol{\Lambda}}\boldsymbol{E}(\boldsymbol{\beta})' \boldsymbol{P}_{\dot{\boldsymbol{z}}} \right) = \text{tr} \left( \boldsymbol{P}_{\dot{\boldsymbol{z}}}\boldsymbol{E}(\boldsymbol{\beta})   \boldsymbol{\Lambda}^{\ast}\boldsymbol{\Lambda}^{\ast \prime}\boldsymbol{E}(\boldsymbol{\beta})' \boldsymbol{P}_{\dot{\boldsymbol{z}}} \right) = \text{tr} \left( \boldsymbol{\Lambda}^{\ast \prime}\boldsymbol{E}(\boldsymbol{\beta})' \boldsymbol{P}_{\dot{\boldsymbol{z}}} \boldsymbol{E}(\boldsymbol{\beta})   \boldsymbol{\Lambda}^{\ast} \right) = \text{tr}\left( \boldsymbol{\Lambda} ^{\ast \prime}\boldsymbol{\Sigma}(\boldsymbol{\beta})\boldsymbol{\Lambda} ^{\ast}\right)$.

\end{document}